# Energy Efficient Distributed Processing for IoT

Barzan A. Yosuf, M. Musa, Taisir Elgorashi, Jaafar Elmirghani (Senior Member, IEEE)

School of Electronic and Electrical Engineering, University of Leeds, Leeds LS2 9JT, U.K.

Corresponding author: Barzan A. Yosuf (elbay@leeds.ac.uk).

This work was supported in part by the Engineering and Physical Sciences Research Council (EPSRC), in part by the INTERNET project under Grant EP/H040536/1, and in part by the STAR projects under Grant EP/K016873/1 and in part by the TOWS project under Grant EP/S016570/1.

**ABSTRACT** The number of connected objects in the Internet of Things (IoT) is growing exponentially. IoT devices are expected to be between 26 billion to 50 billion devices in the near future. This figure can grow even further due to the production of miniaturized portable devices that are light weight, energy and cost efficient together with the widespread use of the Internet and the added value organizations and individuals can gain from IoT devices. The IoT objects' data is traditionally transported to distant clouds via the core network for processing purposes however, this can be prohibitively costly in terms of energy consumption. In this paper, the entire IoT-fog-cloud architecture is modelled, the service placement problem is optimized through Mixed Integer Linear Programming (MILP) and the total power consumption is jointly minimized for processing and networking. Four aspects of IoT service placements are examined: 1) non-splittable services, 2) splittable services, 3) inter-service processing overhead for sub-service synchronization and 4) deployment of special-purpose cloud data centers (SP-DCs) . The results showed that for a capacitated problem, service splitting introduces power consumption savings of up to 86% compared to 46% with non-splittable services in relation to processing in general-purpose data centers (GP-DCs). Moreover, it is observed that the inter sub-service processing overhead has a great influence on the total number of service splits. However much insignificant the ratio of the processing overhead, the results showed that this is not a trivial matter and hence much attention needs to paid to this area in order to make the best use of the resources that are available in the edge of the network. Moreover, the optimization results showed that, for very high demands, power savings of up to 50% could be achieved with SP-DCs compared to 30% with GP-DCs.

## I. INTRODUCTION

The number of objects connected to the Internet is growing at unprecedented rates leading to the concept of Internet of Things (IoT) [1],[2]. In 2011, this number surpassed the world's population and beyond 2020, interconnected devices are expected to range between 26 billion to 50 billion devices [3], [4]. This increase is directly related to the technological advancement in the past decades that enabled the production of miniaturized portable devices that are light weight, energy and cost efficient together with the widespread use of the Internet and the added value organizations and individuals can gain from IoT devices. Multitudes of IoT applications have the potential to transform all aspects of life, some already exist while others are yet to be realized. The massive amounts of data produced if processed centrally by conventional clouds would lead to slow decision making and increased pressure on the already overwhelmed network. Autonomous vehicles for example are reported to generate data that is in the range of 1 GB per second [5]. It is evidently clear that transporting all of this data to the cloud for processing is prohibitively costly in terms of bandwidth requirements and energy efficiency [6]. In the past, the main focus of Information and Communication Technologies (ICT) was primarily fixated on performance only. Little or no attention was paid to the amount of power ICT based components consumed and consequently their adverse impact on our environment. The focus has now shifted towards energy efficiency, due to the rising cost of electricity, resource scarcity and increasing emission of carbon dioxide ($CO_2$) [7]. It is reported that $CO_2$ emissions due to ICT technologies are increasing at an alarming rate of 6% per year. Given this growth rate, the Internet can become responsible for up to 12% of the global emissions by 2030 and cloud data centers which are at the heart of the IoT are one of the major components of ICT [8].

In this direction, distributed processing has been proposed by industry and academia as an effective strategy to curb the pressure imposed by the formidable scale of IoT [9]. Fog computing for instance, is a variant of distributed processing which promises to tackle the aforementioned challenges by utilizing the already available computational, storage, and networking resources for processing of IoT data at the edge of the network [10]. Oftentimes, decision making can be made better and quicker if collected data is processed closer to the source [11]. Currently, fog computing is still in its infancy and a standardized architecture has yet to be agreed. Thus alternative IoT architectures are increasingly being studied in the research community in terms of efficient resource management and the interplay between the edge devices (fog) and the core (cloud), since fog is regarded as a powerful complement to the cloud [12]. A proper resource management scheme is crucial in the fog, as services can be placed in a highly energy inefficient server or even further from the source node which results in higher communication latency [6]. Since contemporary fog devices have limited processing, storage and communication capabilities, offloading services to more resourceful clouds or even fogs becomes a necessity [13]. It is expected that through cooperation between fogs and the centralized cloud, a more efficient and greener computing platform can be achieved [14].

Generally, in a fog architecture, large number of devices exist at the edge of the network, which collectively provide enormous amounts of computational power, that, if used, may help in curbing the unnecessary data exchange between the IoT and the centralized cloud [15]. These devices are heterogeneous in nature in terms of resources. This poses a number of challenges in the optimum design of architectures, protocols and hardware of future IoT based networks. Hence, proper resource management and network design frameworks are needed [16]. These should take into account important dimensions such as but not limited to energy efficiency, due to its impact on our environment [17], resilience, due to mission critical services [4], [17], [18] and end-device cooperation due to traffic bifurcations which lead to inter-service communication [9], [19]. Fog based solutions have been proposed to improve various performance metrics in terms of energy, latency, QoS, etc through various approaches such as resource allocation [12], [26]-[30] and architectural design and planning [21], [24], [25]. The reader is referred to the works in [17] and [20], for architectural design imperatives of fog networks and a detailed taxonomy of fog based solutions, respectively.

It is observed that each of the approaches proposed in all of the aforementioned studies does not consider fog solutions that offer network designers insight into energy efficiency in short-term (capacitated) and long-term (un-capacitated) optical based fog networks. Moreover, our previous works considered energy efficient solutions in cloud and core networks, IoT and mobile networks using MILP techniques considering a variety of scenarios including big data processing in core networks [21], [22], design of energy efficient optical architectures [23]–[25], and data centers [26], content distribution [27] and caching [28], network coding [29], NFV and big data in mobile networks [30], [31] and virtualization and process embedding in IoT based networks [32], [33].

In contrast, the work in this paper aims first to model the entire IoT infrastructure in which all layers of the networking domains such as end devices, access, metro and core are taken into account from the moment an IoT service is launched until it is hosted on the ultimate destination which is the cloud DC, accessed via the core network. A Passive Optical Network (PON) has been proposed to support the fog infrastructure in the access domain as it is increasingly utilized due to its suitability for data intensive applications as they provide high bit rates, relatively low cost and high scalability [6]. An Ethernet based network is considered in the metro to aggregate traffic from the PON towards the cloud DCs in the core domain. An IP/WDM core network is considered to provide access to cloud DCs, in which a large number of servers are inter-connected via a LAN network.

One of our main contributions in this work is the inclusion of the optical core network to provide access to the Cloud DC which is currently not supported by any of the aforementioned studies. Furthermore, several design characteristics that affect the power consumption of the fog approach are considered. Those include 1) granular power consumption profile of networking and processing devices, 2) Power Usage Effectiveness (PUE) to account for cooling [34] requirement in higher capacity devices found in the access, metro, core and cloud layers, 3) service splitting and the prospect of improved server packing in the fog layers, 4) deployment of special purpose DCs (SP-DCs) in the core network in addition to its general purpose DC (GP-DC) counterpart, and 4) inter-service processing overhead to account for synchronization between sub-services.

The remainder of this paper is organized as follows. Section II discusses the related work. Section III presents the proposed architecture followed by the MILP formulation of the service distribution optimization problem. Section V presents and discusses the performance evaluation of the proposed model which includes comparable studies of non-splittable services vs. splittable ones. The fog approach is evaluated in both un-capacitated and capacitated network design cases for services with no splitting. Furthermore, the impact of service splitting on improving the power consumption in a capacitated case is studied. Also, the performance of the fog approach is further examined by deploying SP-DC in both the capacitated and un-capacitated cases. Section VI focuses on the impact of processing overhead due to synchronization traffic between IoT sub-services. Section VII concludes the contributions of the paper.

## II. RELATED WORK

The problem of efficient resource provisioning in distributed architectures such as the fog, has been investigated extensively from different perspectives including energy efficiency, latency, QoS, resilience, handovers, etc. The focus in the literature has shifted towards making the whole IoT infrastructure energy efficient [35] as opposed to optimizing only individual layers namely the device layer, access layer or the cloud. The works in [36] and [37] proposed the use of PONs to extend cloud and fog services closer to the user premises, respectively. Optical based networks are expected to become increasingly important to support edge and fog computing in the next decades. Although no particular algorithmic or optimization model was proposed but nevertheless detailed discussions were provided on how the architecture in question can improve QoS and how different distributed fog resources located in the user premises can efficiently be managed. The authors of [35] proposed an energy efficient IoT architecture in which sensors' sleep intervals are predicted based on their remaining battery level and as a result resources of the cloud can be better utilized by re-provisioning them when the sensory nodes are in sleep mode. The main contribution of the work is centered around developing a mechanism to predict the sleep intervals of sensor nodes based upon certain sensor variables such as battery level and previous usage history.

The work in [38] mathematically models the entire fog network from the end terminals (TNs) to the cloud data centers located in the core network. The TN nodes sense data and transmit the same to the fog tiers, either to be processed by fog nodes or to be forwarded to the cloud for further analysis. The performance of the fog approach in provisioning for IoT applications is investigated by considering several dimensions such as power consumption, $CO_2$ emissions and service latencies in the fog network compared to the baseline cloud system. Their results indicate that the fog computing approach is only beneficial when there is high number of latency-sensitive applications. Although fog computing was comprehensively studied, the authors made no mention of the practical networking or processing hardware that were used in obtaining their results. In another work, the authors of [39] compare the efficiencies of highly distributed edge devices called nano data centers that can host and distribute user contents in a P2P fashion. These edge servers are comprised of Raspberry Pi's that are low power single board computers. The authors investigate the system performance through a time based and a flow based power consumption model. For devices that are highly shared by many users and services, the authors adopt a flow-based model whilst a time based model is used for equipment that are close to end-users.

The work of [40] proposes a framework for cloudlet based network design and planning. The focus of the work is primarily centered around designing a network based on TDM-PON to optimize the network infrastructure cost whilst meeting latency constraints only. The problem is formulated as a Mixed Integer Non Linear Programming (MINLP) model which is utilized to identify efficient cloudlet placement locations and optimal assignment of ONUs to cloudlets. The feasibility of the proposed model is evaluated against urban, suburban and rural scenarios, which provide guidance on the installation and maintenance costs. In another work [41], a generic fiber-wireless architecture is proposed which supports coexistence of the centralized cloud and distributed mobile edge computing (MEC) for IoT connectivity. A distributed game theoretic algorithm is developed to support collaborative computational offloading between the cloud and MEC. Numerical results show very low energy consumption is achieved compared to the baseline which is the optimal case that cannot be realized in practice, hence the distributed approach is used to reduce complexities. The authors of [42] put forth a capacity planning framework that improves the resource utilization of a hierarchical edge cloud network whilst simultaneously meeting QoS requirements in terms of response delay. They do this, by taking advantage of diverse demands for CPU, GPU and network resources.

The authors of [43] formulate the service distribution problem in an IoT-Cloud architecture using a linear program whose solution results in the optimum placement of IoT service functions and the routing of network flows across a multi-layer architecture consisting of devices, access and cloud layers. The total energy consumption is minimized whilst meeting the end-user latency demands. In another work [44], the service allocation problem is formulated as an integer programming optimization, whose objective function is to minimize the total latency experienced by IoT services, subject to capacity constraints at the various layers of the proposed fog architecture. The IoT service requests are considered to be generic, ranging between 10 – 50 homogenous requests. The delay is minimized by placing the less demanding services as close as possible to the IoT devices whilst the medium and high demanding services are placed higher up the fog network. In their work, IoT devices have been excluded from hosting any type of data processing.

Similarly, the authors in [45] propose a generic algorithmic for the placement of IoT services in a fog-cloud framework. The IoT services are considered as multiple modules that are collectively used to deliver a full application. A specific algorithm is used to efficiently deploy application modules dynamically across the fog-cloud infrastructure close to the source terminals in the fog layer. The performance of the proposed solution is addressed through evaluation of latency and efficient resource utilization and it is claimed that it can be extended to include further design dimensions. In [46], an Integer Linear Program (ILP) is proposed to model the problem of resource provisioning from the perspective of service providers, in the context of the heterogeneous Internet of Things, where the objective function is to minimize the total monetary costs subject to capacity and latency budgets. The heterogeneity of IoT is modelled through unique profiling of applications and as such 4 different types of applications are considered. The topology considered comprises of a Metropolitan Area Network (MAN) and consists of two hierarchical levels of interconnected rings. The results indicated that the total operational cost is directly impacted by the application computational complexity, compression factor, and latency budget, coupled with proportions of local traffic versus global traffic. The authors in [12] put forth a convex optimization model that addresses the delay-power trade-off in a cloud-fog architecture which consists of four subsystems. The work demonstrated that compromising modestly on computational resources in order to save communication bandwidth and reduce transmission latency, fog computing platforms can significantly complement the performance of cloud computing. The proposed work has given no consideration to the impact of local computation using the devices in the IoT layer.

The authors of [47], unlike the previous aforementioned works, model the IoT service placement in a practical test bed using an ILP formulation by considering several objective functions that address service latency, service migrations and energy efficiency. The optimization model is executed iteratively to allow for the retention of the objective values of previously executed models, thus, the feasibility region continuously decreases since iterations must satisfy previous results. The approach is generic and can be adapted to other resource placement problems. Their results show that for real-time services, latency becomes important and thus services are processed on the nearest fog, while the latency tolerant services can be offloaded to the distant cloud as energy consumption becomes the priority.

## III. THE PROPOSED DISTRIBUTED PROCESSING ARCHITECTURE

We begin by introducing and describing each layer of the proposed architecture depicted in Figure 2. It comprises of four main layers, namely the IoT Devices, Access Fog (AF), Metro Fog (MF) and the Cloud DC (DC). The following subsections will provide further details on the aforementioned layers:

### A. IOT DEVICES (IOT)

The bottom-most layer of the proposed architecture comprises of the IoT devices. These devices are smart, wireless nodes that are used to collect data and transmit the same via the connected access point (AP) to the next layer for processing and analysis, if local resources are insufficient. A WiFi link is considered between the devices and the APs.

### B. CPE FOG (CF)

The Customer Premises Equipment Fog (CF) layer consists of stationary processing units with processing capabilities usually higher than those found in the IoT layer [48]. The processing devices are connected to ONUs of a PON access network [30]. The ONUs are equipped with internal switches to provide connectivity to the CF servers and higher layers of the architecture. Small organizations or even end-users can deploy their own fog infrastructures at locations such as APs, routers, gateways and etc.

### C. ACCESS FOG (AF)

The third layer is still part of the PON access network, however, it differs in terms of processing capability. A number of high-end servers are used to form a fog collocated with the OLT [27], [48]. Thus, a substantial amount of service demands aggregated from the ONUs can be hosted and processed on the fog connected to the OLT Ethernet input. However, the number of servers that can be collocated with the OLT is constrained by space limitations and therefore service demands may need to be relayed to the next layer for processing.

### D. METRO FOG (MF)

The metro network consists of a high-capacity Ethernet switch and a couple of edge routers that act as a gateway to the cloud data centers via the core network. The computational resources available to the metro fog are substantially higher in comparison to the lower fog layer due to the number of users and services it supports, however it still is incomparable to the cloud DC [49].

### E. CLOUD DC (DC)

The cloud layer comprises of a set of data centers that are accessed via the core network. The core network uses IP/WDM technology and it consists of two layers, the IP layer and the optical layer. In the IP layer, an IP core router is deployed at each node to aggregate network traffic from the metro routers. The optical layer is used to interconnect the IP core routers through optical switches and IP/WDM technologies such as EDFAs, transponders and regenerators. Two types of data centers are considered: 1) a general purpose data center (GP-DC), and 2) a special purpose data center (SP-DC). Both data centers are a single hop away from the aggregation core router. As depicted in Figure 1, the local area network (LAN) elements inside both data centers consist of an edge router and a set of high-speed switches to interconnect thousands of servers. Motivated by the sheer computational power of Graphical Processing Units (GPUs) as well as the breakthrough performances in terms of power consumption efficiencies for visual based deep learning algorithms, it is of interest to investigate the impact of deploying such servers inside DCs connected to the core. A SP-DC only performs a specific service i.e. visual processing. On the contrary, the general-purpose data center (GP-DC) is designed to execute a range of generic services, hence, not as power efficient as the SP-DC. NVidia being a leading manufacturer, have reported GPUs to be at least 10 times more efficient than CPUs.

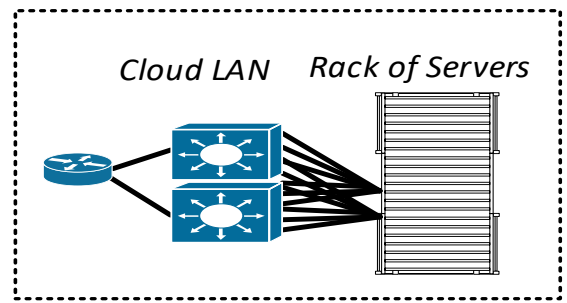

Figure 1 Cloud DC LAN network

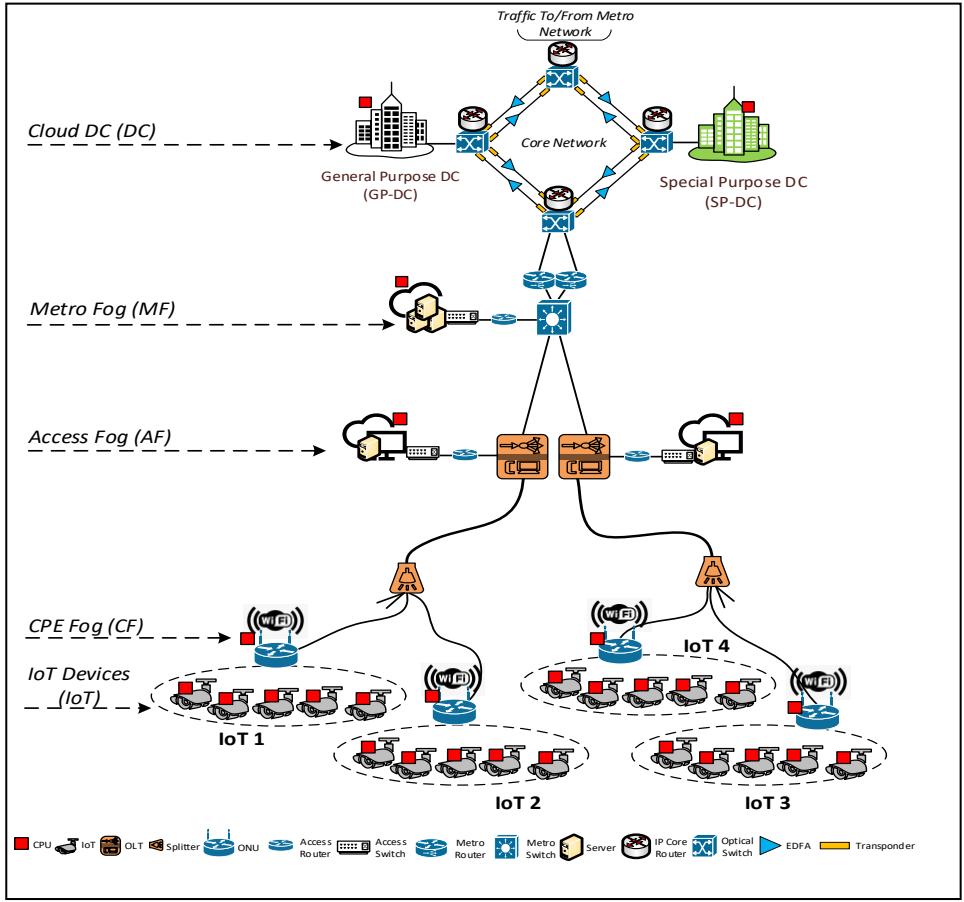

Figure 2 PON-based architecture supported by fog and cloud Infrastructures.

## IV. MILP MODEL

In this section, we develop a MILP model to minimize the total power consumption of the proposed IoT distributed processing architecture shown in Figure 2, by optimizing the placement of IoT service demands. Each demand is characterized by a tuple $D(CPU, BW)$, where CPU is the amount of processing requirement in Million Instructions Per Second (MIPS) and $BW$ is the traffic requirement in bps. The IoT network is modelled as a graph $G(N, L)$, where $N$ is the set of all nodes in the network and $L$ is the set of bidirectional links connecting those nodes together. A subset $I \subset N$ represents the set of the IoT devices in the considered network, whilst a subset $S \subseteq I$ acts as source nodes. A subset of nodes, where $P \subset N$ acts as processing nodes. The processing node $d \in P$ has a maximum computational capacity $\left(C_d^{(CPU)}\right)$ measured in MIPS. Also, each link $(m, n)$, where $m \in N$ and $n \in N_m$, has a maximum bit rate (BR) measured in bps. Before introducing the model, we define the sets, parameters and variables used, as follows:

**Sets:**
$N$ — Set of all nodes.
$N_m$ — Set of neighbor nodes of node m in the proposed architecture
$C$ — Set of core nodes in the IP/WDM network, where $C \subset N$.
$O$ — Set of ONUs in the PON network, where $O \subset N$.
$OT$ — Set of OLTs in the PON network, where $OT \subset N$.
$MR$ — Set of metro routers, where $MR \subset N$.
$MS$ — Set of metro switches, where $MS \subset N$.
$DC$ — Set of data center nodes, where $DC \subset N$.
$I$ — Set of all IoT devices, where $I \subset N$.
$P$ — Set of nodes with processing devices, where $P \subset N$ and $P = I \cup ONU \cup OLT \cup M^{(Sw)} \cup DC$.
$S$ — Set of IoT devices acting as source nodes where $S \subset I$.

**Core Network Parameters:**
$Pe$ — Maximum power consumption of an EDFA in the core network.
$Po$ — Maximum power consumption of an optical switch in the core network.
$Prg$ — Maximum power consumption of a regenerator in the core network.
$Ir$ — Idle power consumption of an IP router port in the core network.
$It$ — Idle power consumption of a transponder in the core network.
$Ie$ — Idle power consumption of an EDFA in the core network.
$Io$ — Idle power consumption of an optical switch in the core network.
$Irg$ — Idle power consumption of a regenerator in the core network.
$B$ — Maximum bit rate of single wavelength.
$W$ — Number of wavelengths in a fibre in the core network.
$\epsilon^{(r)}$ — Energy per bit of a router port, where $\epsilon^{(t)} = \left(\frac{Pt-It}{B}\right)$.
$\epsilon^{(t)}$ — Energy per bit of a transponder, where $\epsilon^{(r)} = \left(\frac{Pr-Ir}{B}\right)$.
$\epsilon^{(e)}$ — Energy per bit of the EDFAs, where $\epsilon^{(e)} = \left(\frac{Pe-Ie}{B}\right)$.
$\epsilon^{(o)}$ — Energy per bit of the optical switches, where $\epsilon^{(o)} = \left(\frac{Po-Io}{B}\right)$.
$\epsilon^{(rg)}$ — Energy per bit of regenerators, where $\epsilon^{(rg)} = \left(\frac{Prg-Irg}{B}\right)$.
$D_{mn}$ — Distance between two core nodes $m$ and $n$, where $m, n \in C$.

| | |
|---|---|
| $Se$ | Span distance between two EDFAs. |
| $Sg$ | Span distance between two regenerators. |
| $A_{mn}$ | Number of EDFAs used on each fiber in the core network from node $m \in C$ to $n \in C$, $A_{mn} = \left\lfloor \left(\frac{D_{mn}}{Se}\right) - 1 \right\rfloor + 2$. |
| $R_{mn}$ | Number of regenerators used between core node $m \in C$ and core node $n \in C$, $R_{mn} = \left\lfloor \left(\frac{D_{mn}}{Sg}\right) - 1 \right\rfloor$. |
| $\mathbb{P}c$ | Power Usage Effectiveness of IP/WDM core network node. |

**Cloud Data Center Parameters:**

| | |
|---|---|
| $P_{(sw)}^{(dc)}$ | Maximum power consumption of Cloud DC switch. |
| $I_{(sw)}^{(dc)}$ | Idle power consumption of Cloud DC switch. |
| $B_{(sw)}^{(dc)}$ | Bit rate of Cloud DC switch. |
| $\epsilon_{(sw)}^{(dc)}$ | Cloud DC switch energy per bit, where $\epsilon_{(sw)}^{(dc)} = \left(\frac{P_{(sw)}^{(dc)} - I_{(sw)}^{(dc)}}{B_{(sw)}^{(dc)}}\right)$. |
| $P_{(R)}^{(dc)}$ | Maximum power consumption of Cloud DC router. |
| $I_{(R)}^{(dc)}$ | Idle power consumption of Cloud DC router. |
| $B_{(R)}^{(dc)}$ | Cloud DC router bit rate. |
| $Pr$ | Maximum power consumption of an IP router port in the core network. |
| $Pt$ | Maximum power consumption of a transponder in the core network. |
| $\epsilon_{(R)}^{(dc)}$ | Energy per bit of a Cloud DC router, where $\epsilon_{(R)}^{(dc)} = \left(\frac{P_{(R)}^{(dc)} - I_{(R)}^{(dc)}}{B_{(R)}^{(dc)}}\right)$. |
| $\mathbb{P}d$ | Power Usage Effectiveness of DC node, for processing and networking. |

**Metro Network and Fog Parameters:**

| | |
|---|---|
| $P_{(sw)}^{(m)}$ | Maximum power consumption of a metro switch. |
| $I_{(sw)}^{(m)}$ | Idle power consumption of a metro switch. |
| $B_{(sw)}^{(m)}$ | Bit rate of a metro switch. |
| $\epsilon_{(sw)}^{(m)}$ | Metro switch energy per bit, where $\epsilon_{(sw)}^{(m)} = \left(\frac{P_{(sw)}^{(m)} - I_{(sw)}^{(m)}}{B_{(sw)}^{(m)}}\right)$. |
| $P_{(sw)}^{(mf)}$ | Maximum power consumption of a metro fog switch. |
| $I_{(sw)}^{(mf)}$ | Idle power consumption of a metro fog switch. |
| $B_{(sw)}^{(mf)}$ | Bit rate of a metro fog switch. |
| $\epsilon_{(sw)}^{(mf)}$ | Metro fog switch energy per bit, where $\epsilon_{(sw)}^{(mf)} = \left(\frac{P_{(sw)}^{(mf)} - I_{(sw)}^{(mf)}}{B_{(sw)}^{(mf)}}\right)$. |
| $P_{(R)}^{(m)}$ | Maximum power consumption of a metro router. |
| $I_{(R)}^{(m)}$ | Idle power consumption of a metro router. |
| $B_{(R)}^{(m)}$ | Bit rate of a metro router. |
| $\epsilon_{(R)}^{(m)}$ | Metro router energy per bit, where $\epsilon_{(R)}^{(m)} = \left(\frac{P_{(R)}^{(m)} - I_{(R)}^{(m)}}{B_{(R)}^{(m)}}\right)$. |
| $P_{(R)}^{(mf)}$ | Maximum power consumption of a metro fog router. |
| $I_{(R)}^{(mf)}$ | Idle power consumption of a metro fog router. |
| $B_{(R)}^{(mf)}$ | Bit rate of a metro fog router. |
| $\epsilon_{(R)}^{(mf)}$ | Metro fog router energy per bit, where $\epsilon_{(R)}^{(mf)} = \left(\frac{P_{(R)}^{(mf)} - I_{(R)}^{(mf)}}{B_{(R)}^{(mf)}}\right)$. |
| $\mathbb{P}m$ | Power Usage Effectiveness of a metro node, for processing and networking. |
| $\mathcal{R}$ | Metro router port redundancy. |

**Access Network and Fog Parameters:**

| | |
|---|---|
| $P^{(ot)}$ | Maximum power consumption of OLT in the PON network. |
| $I^{(ot)}$ | Idle power consumption of OLT in the PON network. |
| $B^{(ot)}$ | Bit rate of OLT in the PON network. |
| $\epsilon^{(ot)}$ | OLT router energy per bit, where $\epsilon^{(ot)} = \left(\frac{P^{(ot)} - I^{(ot)}}{B^{(ot)}}\right)$. |
| $P^{(o)}$ | Maximum power consumption of an ONU in the PON network. |
| $I^{(o)}$ | Idle power consumption of an ONU in the PON network. |
| $B^{(o)}$ | Bit rate of the WiFi interface of an ONU in the PON network. |
| $\epsilon^{(o)}$ | ONU energy per bit, where $\epsilon^{(o)} = \left(\frac{P^{(o)} - I^{(o)}}{B^{(o)}}\right)$. |
| $P_{(sw)}^{(af)}$ | Maximum power consumption of an access fog switch. |
| $I_{(sw)}^{(af)}$ | Idle power consumption of an access fog switch. |
| $B_{(sw)}^{(af)}$ | Bit rate of an access fog switch. |
| $\epsilon_{(sw)}^{(af)}$ | Access fog switch energy per bit, where $\epsilon_{(sw)}^{(af)} = \left(\frac{P_{(sw)}^{(af)} - I_{(sw)}^{(af)}}{B_{(sw)}^{(af)}}\right)$. |
| $P_{(R)}^{(af)}$ | Maximum power consumption of an access fog router. |
| $I_{(R)}^{(af)}$ | Idle power consumption of an access fog router. |
| $B_{(R)}^{(af)}$ | Bit rate of an access fog router. |
| $\epsilon_{(R)}^{(af)}$ | Access fog router energy per bit, where $\epsilon_{(R)}^{(af)} = \left(\frac{P_{(R)}^{(af)} - I_{(R)}^{(af)}}{B_{(R)}^{(af)}}\right)$. |
| $P_{(sw)}^{(cf)}$ | Maximum power consumption of CPE fog switch. |
| $I_{(sw)}^{(cf)}$ | Idle power consumption of an CPE fog switch. |
| $B_{(sw)}^{(cf)}$ | Bit rate of a CPE fog switch. |
| $\epsilon_{(sw)}^{(cf)}$ | CPE fog switch energy per bit, where $\epsilon_{(sw)}^{(cf)} = \left(\frac{P_{(sw)}^{(cf)} - I_{(sw)}^{(cf)}}{B_{(sw)}^{(af)}}\right)$. |
| $\mathbb{P}a$ | Power Usage Effectiveness of an access fog node, for processing and networking. |

**IoT Devices' Parameters:**

| | |
|---|---|
| $P^{(it)}$ | Maximum power consumption of an IoT transceiver. |
| $I^{(iot)}$ | Idle power consumption of an IoT transceiver. |
| $B^{(iot)}$ | Bit rate of the WiFi interface of an IoT device. |
| $\epsilon^{(iot)}$ | IoT WiFi interface energy per bit, where $\epsilon^{(iot)} = \left(\frac{P^{(iot)} - I^{(iot)}}{B^{(iot)}}\right)$. |

**Processing Devices' Parameters:**

| | |
|---|---|
| $P_d^{(pr)}$ | Maximum power consumption of processing device $d \in P$, in Watts. |
| $I_d^{(pr)}$ | Idle power consumption of processing device $d \in P$, in Watts. |
| $C_d^{(pr)}$ | Maximum capacity of processing device $d \in P$ in Million Instructions Per Second (MIPS). |
| $E_d^{(i)}$ | Energy per instruction of processing device $d \in P$, where $E_d^{(i)} = \left(\frac{P_d^{(pr)} - I_d^{(pr)}}{C_d^{(pr)}}\right)$. |

**Application Parameters:**

| | |
|---|---|
| $\delta$ | Portion of the idle power of equipment attributed to the application. |
| $K$ | Number of sub-services an IoT service can be divided into. |
| $\Delta$ | Number of MIPS required to process 1 Mb of traffic. |
| $M$ | Large enough number. |

**Variables:**

| | |
|---|---|
| $\lambda^{sd}$ | Traffic demand between IoT source node $s \in S$ and processing device $d \in P$. |
| $\lambda_{mn}^{sd}$ | Traffic flow between IoT source node $s \in S$ and processing device $d \in P$, traversing link $(m,n)$, where $m \in N, n \in N_m$. |
| $\lambda_d$ | Volume of traffic aggregated by node $d \in N$. |
| $\mathcal{B}_m$ | $\mathcal{B}_m = 1$, if network node $m \in N$ is activated, otherwise $\mathcal{B}_m = 0$. |
| $\theta_d$ | Traffic in node $d \in P$ for processing, where $\theta_d = \lambda_d \Omega_d$. |
| $\Gamma_{mn}$ | $\Gamma_{mn} = 1$, if core network link $m,n$, where $m \in C, n \in (N_m \cap C)$ is activated, otherwise $\Gamma_{mn} = 0$. |
| $\rho^{sd}$ | Processing demand of IoT source node $s \in S$ hosted at processing device $d \in P$. |
| $\Omega^{sd}$ | $\Omega^{sd} = 1$, if processing demand of IoT source node $s \in S$ is hosted at destination node $d \in P$, otherwise $\Omega^{sd} = 0$. |
| $\Omega^d$ | $\Omega^d = 1$, if processing node $d \in P$ is activated, otherwise $\Omega^d = 0$. |
| $\mathcal{N}_d$ | Number of processing servers activated at node $d \in P$. |
| $W_{mn}$ | Number of wavelengths used in fiber link $(m,n)$ in the core network, where link $m,n \in C$. |
| $F_{mn}$ | Number of fibers used on link $m,n \in C$. |
| $Ag_m$ | Number of aggregation router ports activated at IP node $m \in C$. |

The total power consumption of the fog architecture depicted in Figure 2 is composed of three parts: A) Network Power Consumption, B) Processing Power Consumption and C) Intra Processing Node Networks' Power Consumption.

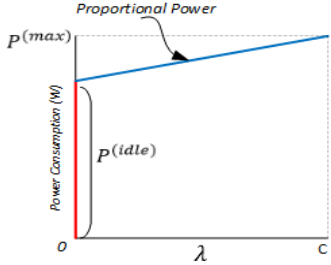

Figure 3 Linear power profile with idle part.

While it is valid to assume that, the desirable power consumption profile should be fully load proportional, however in practical circumstances, this is not the case. It is reported in [50], that almost all devices adopt a linear power profile that consists of an idle and proportional part as depicted in Figure 3. With the former, power is consumed as soon as the device is activated regardless of the load however the latter dependents on various parameters such as frequency, voltage, or workload. In practice, idle power draws a large proportion of the maximum power of a networking/ processing device and hence it cannot be ignored. Since the devices involved in the considered architecture span multiple heterogeneous layers, it becomes a necessity to fairly represent the utilization characteristics of these devices. For example, high-capacity networking elements such as OLTs, metro/core routers and switches are used by many other types of applications in addition to the IoT and it would not make a fair evaluation if the total idle power consumption of these devices were wholly attributed to a small number of IoT services. The total power consumption ($TPC$) considering the linear profile with idle power consumption of a networking or processing device is calculated using equation (1):

$$TPC = \left(\frac{Pmax - Pidle}{C}\right)\lambda + Pidle \quad (1)$$

where $Pidle$ is the idle power consumption of the device which is consumed as soon as the device is activated regardless of the load $\lambda$ and ($Pmax$) is the maximum power consumption of the device, when it is 100% utilised at full capacity C (either in bps or MIPS). The proportional section of the power profile model for networking devices is expressed as energy per bit and likewise, for processing, it is expressed as energy per instruction.

### A. Network Power Consumption ($Net\_PC$):

Under the non-bypass light path approach [51], the core network's IP/WDM total power consumption is composed of:

The power consumption of router ports:

$$\mathbb{P}c\left[\sum_{m\in C}(\epsilon^{(r)}\lambda_m) + \sum_{m\in C}\left(\delta Ir\left(Ag_m + \sum_{n\in(N_m\cap C)}W_{mn}\right)\right)\right] \quad (2)$$

The power consumption of transponders:

$$\mathbb{P}c\left[\sum_{m\in C}(\epsilon^{(t)}\lambda_m) + \sum_{m\in C}\sum_{n\in(N_m\cap C)}(\delta It W_{mn})\right] \quad (3)$$

The power consumption of EDFAs:

$$\mathbb{P}c\left[\sum_{m\in C}(\epsilon^{(t)}\lambda_m A_{mn}F_{mn}) + \sum_{m\in C}\sum_{n\in(N_m\cap C)}(\delta Ie A_{mn}F_{mn})\right] \quad (4)$$

The power consumption of optical switches:

$$\mathbb{P}c\left[\sum_{m\in C}(\epsilon^{(o)}\lambda_m) + \sum_{m\in C}(\delta Io\mathcal{B}_m)\right] \quad (5)$$

The power consumption of regenerators:

$$\mathbb{P}c\left[\sum_{m\in C}(\epsilon^{(rg)}\lambda_m Rg_{mn}W_{mn}) + \sum_{m\in C}\sum_{n\in(N_m\cap C)}(Irg\, Rg_{mn}W_{mn})\right] \quad (6)$$

The metro network's power consumption consists of the power consumption of metro routers and switch, which is given as:

$$\mathbb{P}m\left[\mathcal{R}\sum_{m\in MR}(\epsilon^{(m)}_{(R)}\lambda_m) + \mathcal{R}\sum_{m\in MR}(\delta I^{(m)}_{(R)}\mathcal{B}_m) + \sum_{m\in MS}(\epsilon^{(m)}_{(sw)}\lambda_m) + \sum_{m\in MS}(\delta I^{(m)}_{(sw)}\mathcal{B}_m)\right] \quad (7)$$

The access network's power consumption consists of the power consumption of OLT and ONU devices, which is given as:

$$\mathbb{P}a\left[\sum_{m\in OT}(\epsilon^{(ot)}\lambda_m) + \sum_{m\in OLT}(\delta I^{(ot)}\mathcal{B}_m) + \sum_{m\in O}(\epsilon^{(o)}\lambda_m) + \sum_{m\in O}(\delta I^{(ot)}\mathcal{B}_m)\right] \quad (8)$$

The IoT devices' communication interfaces power consumption is given as:

$$\sum_{m\in I}(\epsilon^{(iot)}\lambda_m) + \sum_{m\in I}(\delta I^{(iot)}\mathcal{B}_m) \quad (9)$$

### B. Processing Power Consumption ($Pr\_PC$):

The total power consumption of the processing devices (or servers) is composed of:

The processing power consumption of IoT devices:

$$\sum_{s\in S}\sum_{d\in I}(E^{(i)}_d \rho^{sd}) + \sum_{d\in I}(I^{(pr)}\mathcal{N}_d) \quad (10)$$

The processing power consumption of CPE fog (CF) servers:

$$\sum_{s\in S}\sum_{d\in O}(E^{(i)}_d \rho^{sd}) + \sum_{d\in O}(I^{(pr)}_d \mathcal{N}_d) \quad (11)$$

The processing power consumption of access fog (AF) servers:

$$\mathbb{P}a\left[\sum_{s\in S}\sum_{d\in OT}(E^{(i)}_d \rho^{sd}) + \sum_{d\in OT}I^{(pr)}_d \mathcal{N}_d\right] \quad (12)$$

The processing power consumption of metro fog (MF) servers:

$$\mathbb{P}m\left[\sum_{s\in S}\sum_{d\in MS}(E^{(i)}_d \rho^{sd}) + \sum_{d\in MS}(I^{(pr)}_d \mathcal{N}_d)\right] \quad (13)$$

The processing power consumption of metro fog (MF) servers:

$$\mathbb{P}d \left[ \sum_{s \in S} \sum_{d \in DC} \left( E_d^{(i)} \rho^{sd} \right) + \sum_{d \in DC} \left( I_d^{(pr)} \mathcal{N}_d \right) \right] \quad (14)$$

### C. Network inside Processing Nodes' Power Consumption ($Net\_Pr\_PC$):

The cloud DCs' network power consumption is composed of the power consumption of cloud DC routers and switches:

$$\mathbb{P}d \left[ \sum_{d \in DC} \left( \epsilon_{(R)}^{(dc)} \theta_d \right) + \sum_{d \in DC} \left( \delta I_{(R)}^{(dc)} \Omega^d \right) + \sum_{d \in DC} \left( \epsilon_{(sw)}^{(dc)} \theta_d \right) + \sum_{d \in DC} \left( \delta I_{(sw)}^{(dc)} \Omega^d \right) \right] \quad (15)$$

The metro fog network power consumption consists of power consumption of metro fog routers and switches, which is given as:

$$\mathbb{P}m \left[ \sum_{d \in MS} \left( \epsilon_{(R)}^{(mf)} \theta_d \right) + \sum_{d \in MS} \left( \delta I_{(R)}^{(mf)} \Omega^d \right) + \sum_{d \in MS} \left( \epsilon_{(sw)}^{(mf)} \theta_d \right) + \sum_{d \in MS} \left( \delta I_{(sw)}^{(mf)} \Omega^d \right) \right] \quad (16)$$

The CF network power consumption consists of power consumptions of CF switches which is given as:

$$\sum_{d \in O} \left( \epsilon_{(sw)}^{(cf)} \theta_d \right) + \sum_{d \in O} \left( I_{(sw)}^{(cf)} \Omega^d \right) \quad (17)$$

The MILP model's objective function is to minimize the total power consumption as follows:

**Minimize**: $Net\_PC + Pr\_PC + Net\_Pr\_PC$

**Subject to the following constraints:**

$$\sum_{n \in N_m} \lambda_{mn}^{sd} - \sum_{n \in N_m} \lambda_{nm}^{sd} = \begin{cases} \lambda_{sd} & m = s \\ -\lambda_{sd} & m = d \\ 0 & otherwise \end{cases} \quad (18)$$

$$\forall s \in S, d \in P, m \in N : s \neq d.$$

Constraint (18) conserves traffic from a source node to a destination node in the considered topology depicted in Figure 2. It ensures that the total incoming traffic at a node is equal to the total outgoing traffic of that node; unless the node in question is either the source node or the destination node.

$$\sum_{d \in P} \rho^{sd} = D_s^{(CPU)} \quad \forall s \in S \quad (19)$$

Constraint (19) ensures that processing service per IoT source node $s \in S$ is met at a given destination node.

$$\rho^{sd} \geq \Omega^{sd} \quad \forall s \in S, d \in P \quad (20)$$
$$\rho^{sd} \leq M \Omega^{sd} \quad \forall s \in S, d \in P \quad (21)$$

Constraints (20) and (21) are used in the conversion of $\rho^{sd}$ into its binary equitant. When $\rho^{sd} = 1$, the source node $s \in S$ processes its CPU service request at destination node $d \in P$.

$$\sum_{d \in P} \Omega^{sd} \leq K \quad \forall s \in S \quad (22)$$

Constraint (22) ensures that the number of sub-services a processing demand can be divided into is less than or equal to K, hence $K = 1$ implies no service splitting is allowed.

$$\mathcal{N}_d \leq \mathcal{V}_d \quad \forall d \in P \quad (23)$$

Constraint (23) ensures that the number of servers activated at a processing node $d \in P$, does not exceed the maximum available number of servers in that node.

$$\sum_{s \in I} \Omega^{sd} \geq \Omega^d \quad \forall d \in P \quad (24)$$
$$\sum_{s \in I} \Omega^{sd} \leq M \Omega^d \quad \forall d \in P \quad (25)$$

Constraints (24) and (25) are used to ensure that, the binary variable $\Omega^d = 1$ if processing node $d \in P$ is activated, otherwise $\Omega^d = 0$.

$$\lambda_m = \sum_{\substack{s \in S: \\ m=s}} \sum_{d \in P} \sum_{n \in N_m} \lambda_{mn}^{sd} + \sum_{\substack{s \in S: \\ m \neq s}} \sum_{\substack{d \in P: \\ s \neq d}} \sum_{n \in N_m} \lambda_{nm}^{sd} \quad (26)$$

$$\forall m \in S$$

$$\lambda_m = \sum_{\substack{s \in S: \\ m \neq s}} \sum_{\substack{d \in P: \\ s \neq d}} \sum_{n \in N_m} \lambda_{nm}^{sd} \quad (27)$$

$$\forall m \in (I \cup OLT \cup M^{(Sw)} \cup M^{(R)} \cup DC)$$

$$\lambda_m = \sum_{s \in S} \sum_{\substack{d \in P: \\ s \neq d}} \sum_{\substack{n \in N_m: \\ n \in (N_m \cap C)}} \lambda_{mn}^{sd} \quad \forall m \in C \quad (28)$$

Constraint (26) gives the traffic generated or received by an IoT node with the first term representing its role as a source and the second term representing IoT node serving demands of other IoT nodes. Constraint (27) gives the traffic traversing/ received by a node of the access, metro and cloud network. Constraint (28) gives the traffic traversing the core nodes.

$$\theta_d \leq M \Omega^d \quad \forall d \in P \quad (29)$$
$$\theta_d \leq \lambda_d \quad \forall d \in P \quad (30)$$
$$\theta_d \geq \lambda_d - (1 - \Omega^d)M \quad \forall d \in P \quad (31)$$

Constraints (29), (30) and (31) are used to linearize the non-linear equation $\lambda_d \Omega_d$, where $d \in P$. This ensures that traffic on a processing node $d \in P$ is only accounted for if it is destined to that node for processing.

$$\lambda_m \geq \mathcal{B}_m \quad \forall m \in N \quad (32)$$
$$\lambda_m \leq M \mathcal{B}_m \quad \forall m \in N \quad (33)$$

Constraints (32) and (33) are used to ensure that, the binary variable $\mathcal{B}_m = 1$ if network node $m \in N$ is activated, otherwise $\mathcal{B}_m = 0$.

$$\lambda^{sd} = D_s^{(BW)} \Omega_{sd} \quad \forall s \in S, d \in P \quad (34)$$

Constraint (34) ensures that traffic is only directed to the destination node that is hosting a processing service.

$$\sum_{s \in S} \sum_{\substack{d \in P: \\ s \neq d}} \lambda_{mn}^{sd} \leq C_{mn} \quad (35)$$

$$\forall m \in \left(I \cup ONU \cup OLT \cup M^{(Sw)} \cup M^{(R)} \cup DC\right): n \in N_m$$

Constraint (35) ensures that the total traffic carried on link $m, n$, in the metro and access layer, does not exceed its capacity in Mbps.

$$Ag_m \geq \frac{\lambda_m}{B} \quad \forall m \in C \quad (36)$$

Constraint (36) gives the number of aggregation router ports at each IP/WDM node.

$$\sum_{s \in S} \sum_{\substack{d \in P: \\ s \neq d}} \lambda_{mn}^{sd} \leq W_{mn} B \quad \forall m \in C: n \in (C \cap N_m) \quad (37)$$

$$W_{mn} \leq WF_{mn} \quad \forall m \in C: n \in (C \cap N_m) \quad (38)$$

Constraints (37) and (38) represent the physical link capacity of the IP/WDM optical links. Constraint (37) ensures that the total traffic on a link does not exceed the capacity of a single wavelength while constraint (38) ensures the total number of wavelength channels does not exceed the capacity of a single fiber link.

## V. PERFORMANCE EVALUATION

In order to evaluate the performance of the energy efficient distributed processing model, we use an architecture which consists 20 IoT devices divided into 4 groups uniformly, hence each group is connected to the PON network via a single ONU. We have assumed that the aforementioned devices are "intelligent" in nature in terms of their processing ability, hence service requests can be fulfilled by local CPUs. Table 1 - Table 5 shows the parameters of networking devices, processing servers, intra processing networking devices, PUE values and the core network, respectively.

### A. WORKLOAD DEFINITION

In our evaluations, we have made CPU requirement proportional to traffic (BW), such that, for every bit of traffic 1000 MIPS is required. Although, it is beyond the scope of the work in this paper, measuring CPU efficiency by MIPS is not an accurate benchmark, since different CPUs have different architectures, hence varied performances for the same service. Nevertheless, this does not stop us from making a starting point by consulting the literature in order to obtain realistic values. In [89], the authors have reported that for a specific visual processing algorithm referred to as Analyze Then Compress (ATC), for a file of 10kB, 69.23 MIPS are required for processing for visual object recognition. Thus, through simple calculations we derived how many MIPS are required ($\Delta$) to process 1Mb of traffic as follows, using equation (39):

$$\Delta = \frac{69.23}{0.08} \cong 865.4. \quad (39)$$

For the sake of simplicity and staying conservative, we assume that each 1Mb of traffic requires approximately 1000 MIPS for processing. As for the bandwidth requirement, we used an online tool to estimate the required data rates for different resolutions and this was estimated to be between 1 – 10 Mbps, which covers video resolutions between $1024 \times 720$ to $1600 \times 1200$ at 30 frames per second [52]. The CPU workload intensity is then calculated by multiplying the $\Delta$ by the amount of traffic. Thus, this makes the CPU demand proportional to the size of the traffic due to the assumption that, the higher the traffic, the more features a video file will hold, thus more CPU instructions are required to process that file.

### B. POWER CONSUMPTION DATA

The network data in Table 1 consist of the maximum and idle power consumptions, bit rate and a portion of idle power ($\delta$) if it applies. we have made use of equipment datasheets where possible to report the values, however, it is not always feasible to obtain this information, hence, we make realistic assumptions based on the literature. In terms of idle power consumption, based on [39], most high capacity networking equipment such as metro/core routers and switches consume 90% of the equipment's' maximum power consumption. As for processing servers' idle power consumption, based on [53], we assume it is 60% of the maximum power consumption of the CPU. Moreover, we assume that IoT applications are only responsible for a portion of the maximum idle power. This assumption is valid. For instance, metro switches are used to serve thousands of different users simultaneously, thus it would not make a fair analysis if all of $P_{idle}$ was attributed to a specific application like the one considered in this thesis. Thus, we make use of Cisco's visual networking index for the years 2017-2022 to estimate the total traffic of surveillance type applications like the one considered in this work. It is reported that, globally, 3% of all video traffic in the Internet is due to surveillance services, hence the portion of idle power $\delta$ attributed to the application in question is 3% [54].

| Device | Pmax(W) | Pidle(W) | $\delta$ | BR (Gb/s) |
|---|---|---|---|---|
| IoT (WiFi) | 0.56 [55] | 0.34 [56] | - | 0.1 [55] |
| ONU (WiFi) | 15 [57] | 9 [57] | - | 0.3 [57] |
| OLT | 1940 [58] | 60 [58] | 3% | 8600 [58] |
| Metro Router Port | 30 [59] | 27 | 3% | 40 [59] |
| Metro Ethernet Switch | 470 [60] | 423 | 3% | 600 [60] |
| Metro Router Port Redundancy ($\mathcal{R}$) | 2 [6] | | | |

Table 1 Network devices' data for the MILP model.

The processing devices' input data are summarized in Table 3. In order to estimate the processing capacity of the servers in MIPS, we have made use of a technical benchmark, in which, it is reported that Intel high-end servers process 4 instructions/ cycle (I/C) [61]. Thus, to determine the maximum capacity of a processing device we have used the following

$$IPS = clock \times \frac{I}{C} \quad (40)$$

where $\frac{I}{C}$ is the number of instructions a CPU can execute per clock cycle which is given in GHz. To differentiate between the types CPUs and their efficiencies, we set the $\frac{I}{C}$ of metro fog (MF) server as a reference point. The efficiency of the processing decreases as one moves down the hierarchy (from core to the IoT device). At those layers where multiple servers can be deployed, a networking infrastructure becomes a necessity in order to interconnect the multiple active servers. Hence, we have used routers and switches accordingly to achieve this. We have used realistic values for the processing networking equipment in order to differentiate between the many layers of the proposed architecture in Figure 2. Generally, lower layers have been assigned lower specification devices where applicable, for instance, an L3 metro switch is much more power consuming than an L2 switch at the access. summarizes networking equipment used inside processing nodes.

| Node | Device Model | P(W) | I(W) | GHz | k MIPS | W /MIPS | I/C |
|---|---|---|---|---|---|---|---|
| SP-DC Server | NVidia T4 GPU | 75 [62] | 45 | 1.25 [62] | 1080 | 27$\mu$ | 864 |
| GP-DC Server | Intel Xeon E5-2680 | 130 [63] | 78 | 2.7 [63] | 108 | 481$\mu$ | 5 |
| MF Server | Intel X5675 | 95 [64] | 57 | 3.06 [64] | 73.44 | 517$\mu$ | 4 |
| AF Server | Intel Xeon E5-2420 | 95 [65] | 57 | 1.9 [65] | 34.2 | 1111$\mu$ | 3 |
| CF Server | RPi 3 Model B | 12.5 [66] | 2 | 1.2 [67] | 2.4 | 4375$\mu$ | 2 |
| IoT Device | RPi Zero W | 3.96 [66] | 0.5 | 1 [68] | 1 | 3460$\mu$ | 1 |

Table 3 Processing servers' data.

| Device | Pmax (W) | Pidle (W) | BR (Gb/s) | Eb (W/Gb/s) |
|---|---|---|---|---|
| CF Switch | 1.78W [69] | 0.36[69] | 1.6[69] | 0.89 |
| AF Router | 13W[59] | 11.7 | 40[59] | 0.03 |
| AF Switch | 210W[60] | 189 | 240[60] | 0.08 |
| MF Router | 13W[59] | 11.7 | 40[59] | 0.03 |
| MF Switch | 210W [60] | 189 | 600[60] | 0.04 |
| DC LAN Router | 30[59] | 27 | 40[59] | 0.08 |
| DC LAN Switch | 470[60] | 423 | 600[60] | 0.08 |

Table 4 Intra processing node network devices' data.

## C. POWER USAGE EFFECTIVNESS (PUE)

In our evaluations, PUE is not considered for IoT and ONU devices, as there is generally no cooling requirements for them [70]. The power usage effectiveness (PUE) is the ratio of the total power consumed by a facility (i.e. ISP networks, data centers) to the total power consumption of the equipment within the facility (i.e. servers, switches, routers, etc. In 2018, Google reported that one of their data centers is currently operating at a PUE of 1.15. We make use of a report published in 2016 which estimates the PUE values of various data centers base on "Space Type" [94]. Within the report, it is shown that PUE values progressively decrease with the increase in the "Space Type". Thus, in a similar fashion, we increase PUE progressively in the proposed network architecture since the largest "Space Type" is generally hyper-scale data centers connected to the core network. It is assumed that at the access and metro layers processing and networking equipment has the same PUE. The PUE value of the core network is consistent with one of our previous works, and this is assumed to be 1.5 [71]. Table 2 is a summary of the PUE values used in the model.

| Network Layer | PUE |
|---|---|
| IoT Devices | 1 |
| CPE Fog (CF) | 1 |
| Access Fog ($PUE^{(access)}$) | 1.5 |
| Metro Fog ($PUE^{(metro)}$) | 1.4 |
| Cloud DC ($PUE^{(DC)}$) | 1.12 [72] |
| Core Network (($PUE^{(core)}$)) | 1.5 [71] |

Table 2 PUE values used in the MILP model.

We have considered the DC nodes to be only a single hop from the user traffic and the average distance between two neighboring core nodes is assumed to span 2010 km (estimated using google maps based on the AT&T US network topology) [73]. The power consumption of the core network devices used are consistent with our previous work in [74] and all the parameters are summarized in Table 5.

| | |
|---|---|
| Distance between two neighboring EDFAs ($S^{(EDFA)}$) | 80 (km) [74] |
| Number of wavelengths in a fiber ($W$) | 32 [74] |
| Bitrate of a wavelength ($B$) | 40 Gb/s |
| Distance between two neighboring core nodes $D_{mn}$ | 2500km |
| Maximum power consumption of a router port $Pmax^{(r)}$ | 638 (W) [74] |
| Idle power consumption of a router port $Pidle^{(r)}$ | 574.2 (W) |
| Energy per bit of a router port $Eb^{(r)}$ | 1.6 W/Gb/s |
| Maximum power consumption of a transponder $Pmax^{(t)}$ | 129 (W) [74] |
| Idle power consumption of a transponder $Pidle^{(t)}$ | 116 (W) |
| Energy per bit of a transponder $Eb^{(t)}$ | 0.32 (W/Gb/s) |
| Maximum power consumption of an optical switch $Pmax^{(o)}$ | 85 (W) [74] |
| Idle power consumption of a transponder $Pidle^{(o)}$ | 77 (W) |
| Energy per bit of a transponder $Eb^{(o)}$ | 0.2 (W/Gb/s) |
| Maximum power consumption of an optical switch $Pmax^{(e)}$ | 85 (W) [74] |
| Idle power consumption of a transponder $Pidle^{(e)}$ | 11 (W) |
| Energy per bit of a transponder $Eb^{(e)}$ | 0.02 (W/Gb/s) |
| Maximum power consumption of an optical switch $Pmax^{(rg)}$, reach 2500km | 71.4 (W) [74] |
| Idle power consumption of a transponder $Pidle^{(rg)}$ | 64 (W) |
| Energy per bit of a transponder $Eb^{(rg)}$ | 0.19 (W/Gb/s) |
| Portion of the aggregate idle powers attributed to the application ($\delta$) | 3% [54] |

Table 5 Input data of the core network for the MILP Model

## D. CASE STUDY

In this paper, we consider a visual analysis application for surveillance purposes. In the next generation smart cities, video surveillance is considered an important element due to the fact that distributed cameras on a stretch of a busy road or a shopping center could bring city security onto a higher level, thus providing the public with a strong sense of assurance [75]. China has already implemented a system called Skynet, which consists of a massive network of smart CCTV cameras that have AI incorporated into them and it is claimed to cover the whole of Beijing [76]. Although the project is faced with criticism from the Chinese public due to privacy concerns, it was demonstrated to a BBC journalist from a police control room how the system of networked cameras helped to catch the journalist within 7 minutes after an "escape" was staged [77].Thus, we consider video surveillance applications as it has become clear that their widespread deployment is imminent in the future. The sheer data rates associated with video data being collected by a large scale network of intelligent cameras makes it virtually impractical to transport all of that data to the cloud for processing in order to obtain insights. In a news article published in 2013 by The Telegraph, it is reported that there is one surveillance camera for every 11 people in the UK [78]. Thus, we are motivated to investigate visual processing applications through the concept of fog computing to help reduce the implications of unnecessary data exchange with the centralized cloud by hosting parts or all of the service requests in the distributed layers of the fog framework.

### E. POWER CONSUMPTION EVALUATION

We present by results, the outcomes of the proposed energy efficient distributed MILP model for IoT with non-splittable and splittable services. We approach the service placement optimization problem using two design strategies: 1) capacitated and 2) un-capacitated. It is worthy of mention that, IoT devices are in all cases capacitated in terms of processing only. Note that in our evaluations, the network capacity is always sufficient to carry the traffic. Therefore the 'capacitated restriction' applies to the number of processing servers available at each location in our case. For each design problem, we have evaluated the power consumption by considering four scenarios that capture different distribution of source nodes. Scenario #1 consisted of a single IoT source node generating demands, Scenario #2 consisted of 5 IoT sources nodes within the same group, Scenario #3 consisted of 4 IoT source nodes, 1 per group and Scenario #4 consisted of the other end of the extremes which has all the IoTs (20 devices) generating requests for demands. Furthermore, the impact of deploying SP-DCs is also investigated together with inter-service processing overhead needed for synchronization among sub-services.

#### 1) UN-CAPACITATED DESIGN WITH NON-SPLITTABLE SERVICES

In the un-capacitated design approach, it is assumed that the number of processing devices deployed at each node is unrestricted except in the devices located in the IoT layer due to their limited features. The aim of this approach is to determine for a given demand volume, the optimum resources needed to host a given service if there are no restrictions on the network equipment capacity and no restrictions on the number of servers that can be hosted at each site. The goal is also to determine whether it is the optimal choice to build large numbers of devices at a given location in the proposed architecture. Generally, such design problems occur in medium to long term network design planning [79].

### SCENARIO #1

In this scenario, out of the total 20 IoT devices in the model, we take one end of the extremes and assume that only a single IoT device is active at any time instance and the rest of the IoT devices are in the idle mode. As expected and shown Figure 4, for low workload values such as 1000 MIPS, significant savings (98%) can be achieved compared to the baseline solution, where the baseline solution is a scenario where processing is always carried out at the GP-DC. This is due to the local computational resource of the IoT device, hence, the costly overhead of the network and high idle powers of DC servers are avoided. However, as the workload increases and violates the capacity of the IoT device, we begin to see the intervention of the CF nodes as it is only a single hop from the IoT device. In Figure 5, the general trend in this scenario always favors the activation of additional servers attached to the CF node due to its low idle power consumption compared with the servers located in the upper layers of the fog architecture. Moreover, the results indicate promising power savings of at least 70%, at the extreme end of the workloads (10,000 MIPS).

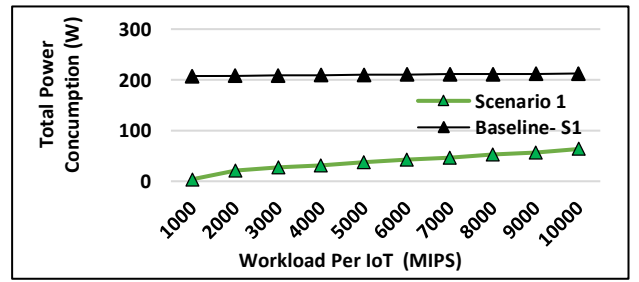

Figure 4 Total power consumption of Scenario #1 using the fog approach vs. the baseline, in the un-capacitated case.

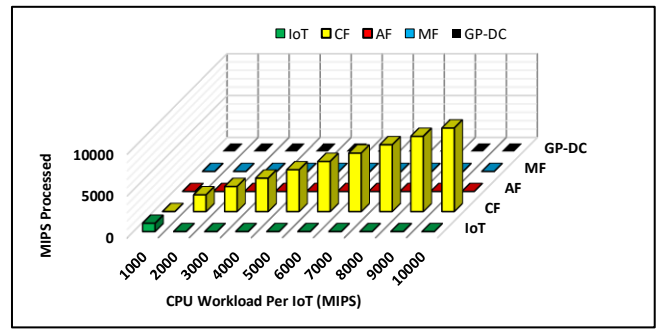

Figure 5 Workload distribution of Scenario #1, using fog, in the un-capacitated case.

## SCENARIO #2

In this scenario, the number of IoT devices demanding computational resources has increased to five devices residing in the same group and connected to the same CF. The trends in this scenario remain the same as scenario #1, except for workload values of 5000 MIPS and beyond. As can be seen in Figure 7, the model decides to allocate all the demands to the metro fog that is connected via the metro network. Although the IoT devices are collocated in the same group and can be allocated to a single CF, the results indicate that activating a large server with a higher idle power and other associated overheads such as networking and PUE, is the optimal choice as multiple servers need to be activated to serve demands of 5000 MIPS and higher at CF. Hence, this gives interesting insights about the potential large scale deployments of such servers at the edge of the network which may not be as energy efficient as larger fog nodes concentrated higher up in the network hierarchy. Although CF servers produce savings of up to 69% for lower ends of the workload but this diminishes as soon as the workload intensity of the services increase and savings drops to 37% as can be seen in Figure 6. With the demands allocated to the MF node, power consumptions savings of up 46% can be achieved, compared to the baseline. As can be seen in Figure 7, the model never utilises the AF server despite its close proximity in terms of distance from the IoT device and the fact that the OLT power consumption is minimal compared to the high capacity Ethernet switch attached to the MF server. The main cause for not choosing to utilise the processing resources of the AF is primarily linked with the high PUE value because the AF and MF are both have identical servers.

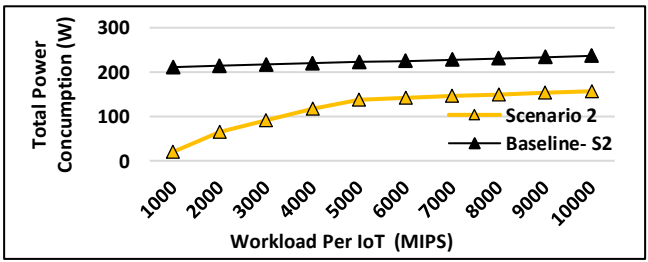

Figure 6 Total power consumption of Scenario #2 using the fog approach vs. the baseline, in the un-capacitated case.

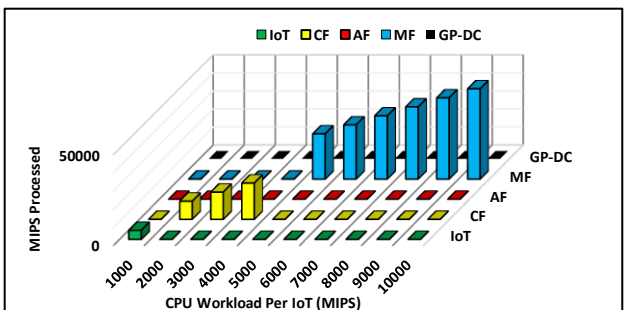

Figure 7 Workload distribution of Scenario #1, using fog, in the un-capacitated case..

## SCENARIO #3

In this scenario, we aim to investigate the effect the location of the IoT devices has on the optimal allocation of services, hence, each request is connected to a separate network. Interestingly, the trends remain unchanged. The results here indicate that for IoT devices located in different parts of the network, activating additional CF servers at the four different locations coupled with the networking overhead at the CF layer ( ONU devices activated) is still the optimal choice as in Scenario #1. For processing demands higher than 5000 MIPS processing moves to the MF as in Scenario #2. The fog approach still produces promising power savings compared to the baseline scenario, as can be seen in Figure 8. When all demands are hosted at the CF layer, savings of up to 66% can be achieved whilst this drops down to 39% when the services are allocated to the MF node.

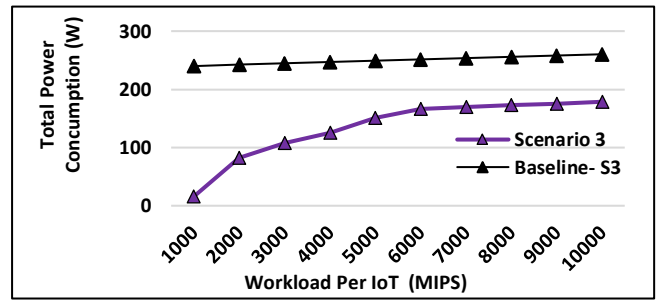

Figure 8 Total power consumption of Scenario #3 using the fog approach vs. the baseline, in the un-capacitated case.

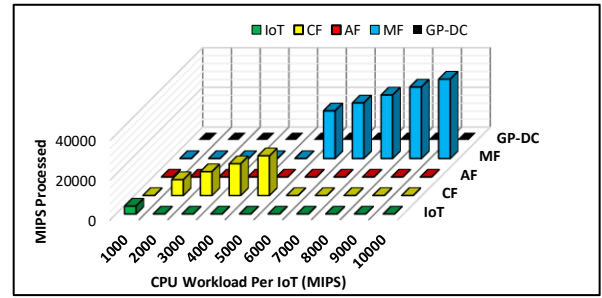

Figure 9 Workload distribution of Scenario #3, using fog, in the un-capacitated case.

## SCENARIO #4

In this scenario, we take the other end of the extremes and assume that all of the IoT devices generate requests for resources simultaneously. With the increase in the number of IoT devices, the volume of demands also increases, hence trends are expected to change. As can be seen in Figure 10, the distributed processing approach still yields total savings of up to 17% at 7000 MIPS, compared to the baseline. However, when the workload volume reaches a certain level, e.g. at 5000 MIPS, the model decides to allocate all of the workload to the centralized cloud data center and bypasses the fog layers all together as serving high demand volumes requires activating multiple servers at the fog layers. This justifies networking overheads and higher idle power associated with activating a large server of improved processing efficiency and PUE at the cloud. At 6000 MIPS and 7000 MIPS the model switches back to the MF as at those particular workload levels, additional servers have to be activated at the cloud, thus the PUE and processing efficiency of the cloud servers do not justify the idle power consumption of multiple cloud servers. Generally, the trends indicate that the optimum processing location is dependent on the number of servers required to process the workload. The only time the CF server is utilized is at workload 4000 MIPS in combination with MF because it saves activating an additional MF server.

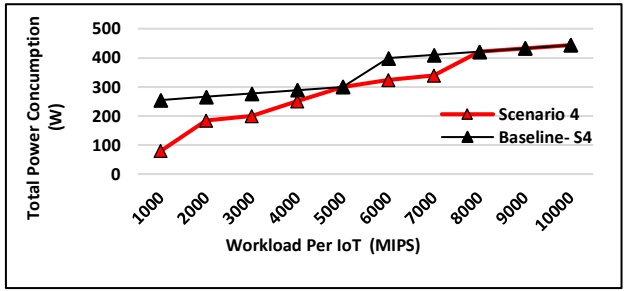

Figure 10 Total power consumption in Scenario #4 of fog vs. baseline, in the un-capacitated case.

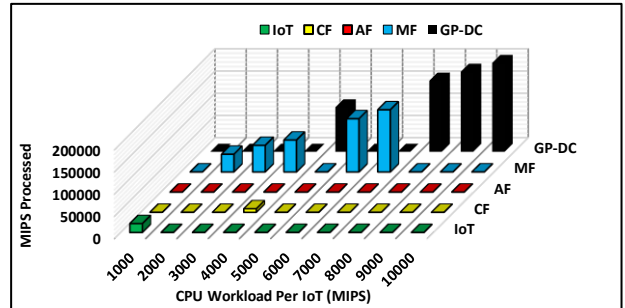

Figure 11 Workload distribution in Scenario #4, using fog, in the un-capacitated case.

2) CAPACITATED DESIGN WITH NON-SPLITTABLE SERVICES

In this subsection, we consider the case where extra capacity cannot be added to the processing nodes in question, hence the problem is capacitated. Such design problems are faced in the short term when the processing nodes are already designed and are in place.

**SCENARIO #1**

In the capacitated design problem, different trends are expected because the prospect of adding extra processing capacity is no longer the case. As can be seen in Figure 13, unlike the trends observed in the scenarios of the un-capacitated case, the AF server is chosen as the next best choice after the IoT local computation and CF capacities have become violated. We have already observed that the AF server is never a good choice in the un-capacitated case and this is primarily down to the high idle power of the servers used inside the AF and the associated PUE required for cooling. Although a bad choice in the longer run, the fog approach still yields savings of up to 46% with AF server as the chosen processing destination, as shown in Figure 12.

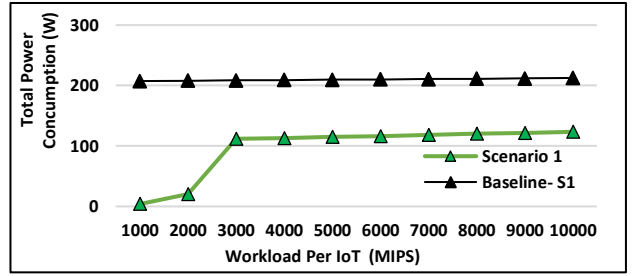

Figure 12 Total power consumption in Scenario #1 of fog vs. baseline, in the capacitated case.

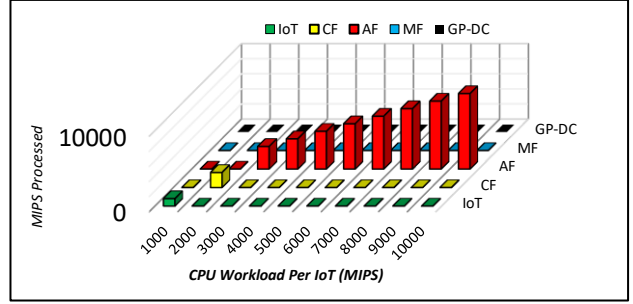

Figure 13 Workload distribution in Scenario #1, using fog, in the capacitated case.

**SCENARIO #2**

In this scenario, we begin to observe the disappearance of the AF node as anticipated due to its lower processing efficiency and higher PUE compared with the MF node, as shown in Figure 15. The total power consumption savings drop down to 41% from 69% for workload volumes of 2000 MIPS in the un-capacitated case. This is mainly the difference between hosting the demands in the CF layer compared to the AF layer. As shown in Figure 14, still a significant amount of power saving is achieved compared to the baseline solution. Although the CF servers had enough capacity to host 9600 MIPS of the total 10,000 MIPS (2000 MIPS/IoT), the model is forced to consolidate processing at the AF layer due to the service splitting constraint forcing processing to take place in a single location because the AF server would need to intervene anyway to process at least 400 MIPS thus packing a single AF server is the optimal choice in this case. This is consistent with previous observations in the un-capacitated case, for lower workload volumes (i.e. 2000 MIPS), the model tends to serve the demands in the lower layers of the fog such as the AF node primarily due to the level of workload since the processing efficiency of the MF server and its lower PUE does not justify the networking overhead for accessing the MF. However, as the workload increases (i.e. 3000 MIPS and higher), the processing efficiency coupled with the lower PUE of the MF server compensates for the networking overhead, hence MF node is chosen as the optimal location to serve the demands as can be seen in Figure 15.

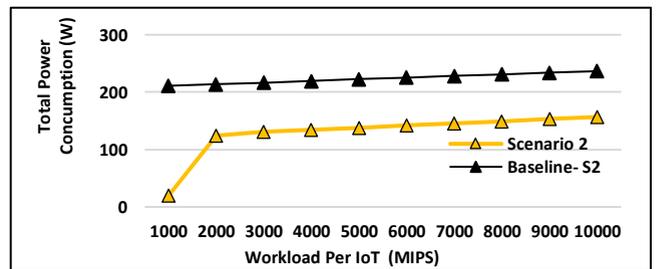

Figure 14 Total power consumption in Scenario #2 of fog vs. baseline, in the capacitated case.

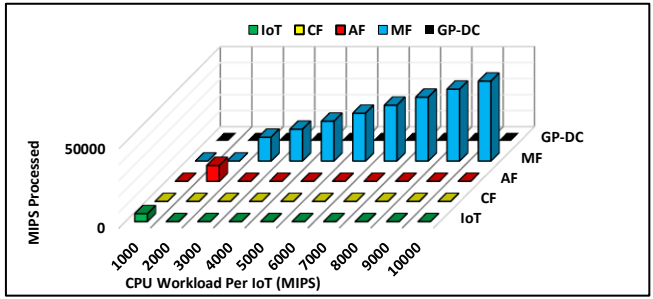

Figure 15 Workload distribution in Scenario #2, using fog, in the capacitated case.

### SCENARIO #3

Figure 19 shows the trends in this scenario is relatively comparable to Scenario #2, except for the case at 2000 MIPS where instead of the AF server, the CF servers are utilized. This is mainly due to the geographical distribution of the IoT source nodes as in this scenario, each CF server has enough capacity to serve its source node and the number of source nodes happen to match the number of CF servers available, hence the high idle power and associated PUE of the higher fog layers like the AF and the MF can be avoided in this case, unlike Scenario #2 at 2000 MIPS. A total saving of up to 66% is achieved at 2000 MIPS and up to 55% saving at higher workloads is achieved, as shown in Figure 16.

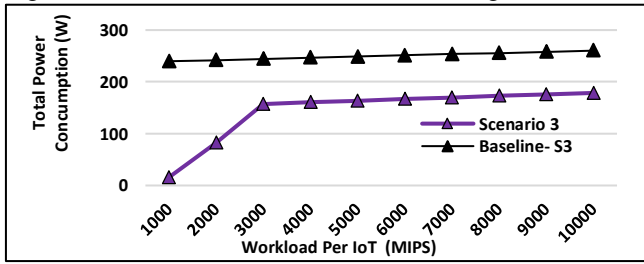

Figure 16 Total power consumption in Scenario #3 of fog vs. baseline, in the capacitated case.

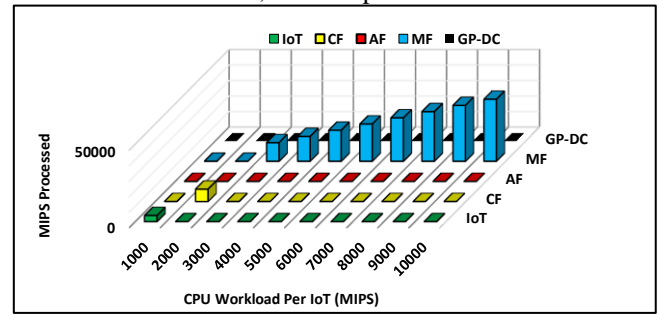

Figure 19 Workload distribution in Scenario #3, using fog, in the capacitated case.

### SCENARIO #4:

In this scenario, we begin to observe the same trends that were found in scenario #4 in the un-capacitated case except that the intervention of the cloud occurs earlier in this scenario at 4000 MIPS. This result proves the consistency of the model since the extra capacity needed to host all the demands at 4000 MIPS, requires multiple servers at the MF node, thus it becomes more efficient to migrate all services to the GP-DC to better pack the already activated servers as it is much more efficient and has a better PUE value. As can be seen in Figure 17, utilization of the MF servers is only beneficial at certain workload values, otherwise once certain number of servers are required, the network overhead to get to the GP-DC justifies the activation of the MF server. Figure 17 shows that there are still substantial savings (about 17%) at 7000 MIPS despite the activation of multiple servers at the MF and its high PUE, compared to the GP-DC.

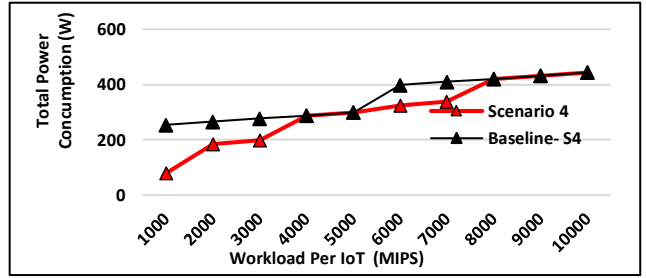

Figure 17 Total power consumption in Scenario #4 of fog vs. baseline, in the capacitated case.

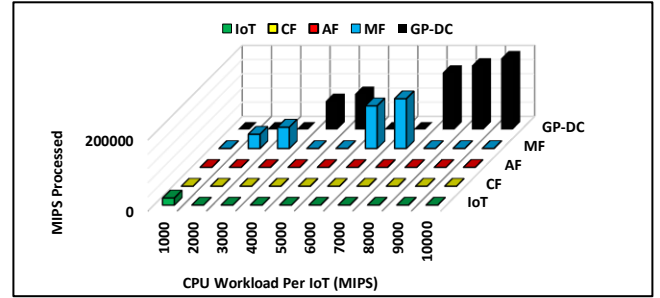

Figure 18 Workload distribution in Scenario #4, using fog, in the capacitated case.

### 3) CAPACITATED DESIGN WITH SPLITTABLE SERVICES

Future IoT services will consist of multiple components, coordinating and communicating over the network to achieve a common service, similar to applications design in Service-Oriented Architectures (SOA) [1]. Each IoT device holds a limited amount of computational resources, given the scale of IoT, each device maybe called on to provide a variety of services [80]. In this direction, this subsection evaluates a scenario in which, processing services can be split into multiple sub-services, hence multiple processing nodes can be utilized to complete a single application service [81]. Figure 20 shows an illustrative example of service splitting whereby an IoT service belongs to a source node which consists of 4 sub-services (S1-S4). A single sub-services (S1) is processed locally whilst the second sub-services (S2) is offloaded via the ONU device to another IoT in the same IoT group. Since the total IoT capacity has been fully utilized, the remaining sub-services (S3 and S4) are processed on the ONU node. It is worthy of mention that this is merely an illustrative example, it does not reflect the optimal distribution of the sub-services in anyway.

The main goal of this subsection is to determine in cases where IoT devices' available CPU capacity is not sufficient to process a service, whether service splitting among numerous processing nodes becomes beneficial in terms of total power reduction, given the added power consumption associated with network overhead after splitting, especially when network equipment have idle power consumption. In our evaluations, service splitting is only said to have occurred if different sub-services of a service are processed in geographically distributed servers, otherwise, if sub-services are all processed on the same processing node, then this is not classed as service splitting mainly due to the same network latency for the sub-services.

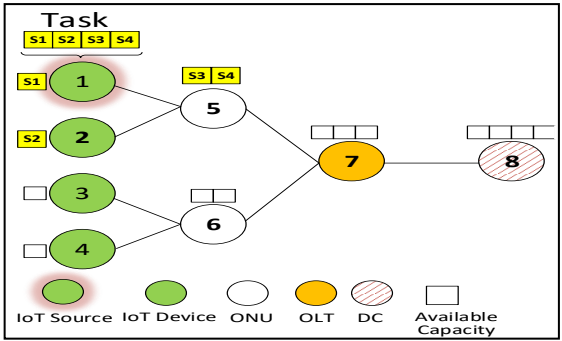

Figure 20 An illustrative example of service splitting in the proposed PON-based architecture.

The MILP model remains unchanged except for a minimal modification to the processing location constraint (22), in order to adapt to the variation introduced by service splitting. Previously, it was assumed that the parameter was $K = 1$, in the current evaluations, $K$ will adopt values from $1 - 5$ to investigate whether service splitting in the short term network design (capacitated), introduces additional savings on top of the fog approach. We also consider the same scenarios as previous evaluations in order to gauge the impact of service splitting on the improvement of the total power consumption. From the previous results, it was found that service splitting is mainly incentivized when processing nodes' capacities are limited hence the current section considers capacitated design problem only.

**SCENARIO #1**

Figure 21 shows, in the case where processing nodes are limited by capacity, with the increase in the number of service splits (i.e. K>1), substantial savings can be made as opposed to the case with no service splits (K=1). The savings are due to the fact that the AF's server idle power is avoided since application services will be processed locally between the IoT and the CF layers, despite the network overhead incurred in getting access to these devices. The total savings achieved by the fog approach with non-splittable services (K=1) was up to 46% compared to the baseline, however this figure increased to 88% when the value of K was changed to 2 (i.e. K=2), as highlighted in Figure 21. Moreover, in Figure 21, when the workload volume increases to 10,000 MIPS, we begin to see a drastic drop in savings. The savings due to service splitting dropped from 88% to about 5% as highlighted in Figure 21. This can be understood by noting that at 10,000 MIPS, 4 CF servers are activated in order to process 9000 MIPS whilst the remaining 1000 MIPS is processed at the IoT source node itself, as shown in Figure 22(b). This was due to the level of workload as the total capacity of the IoT devices in the same group was not enough to host all the workload, hence activating fewer CF servers with relatively higher idle power compensated for activating a large number of IoTs with lower idle power coupled with the associated overhead of networking interfaces. There was an available capacity of 600 MIPS on CF servers but due to processing efficiency and zero networking overhead, the source node was fully packed instead.

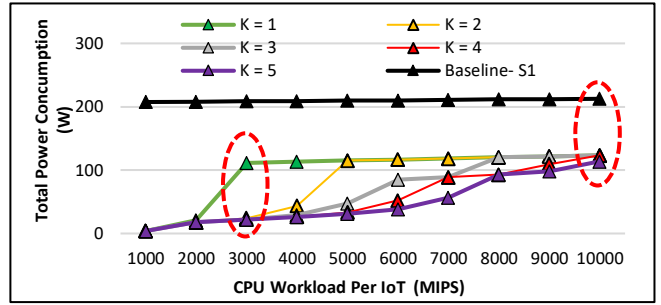

Figure 21 Total power consumption in Scenario #1 using fog, for a range of values of K, in the capacitated case.

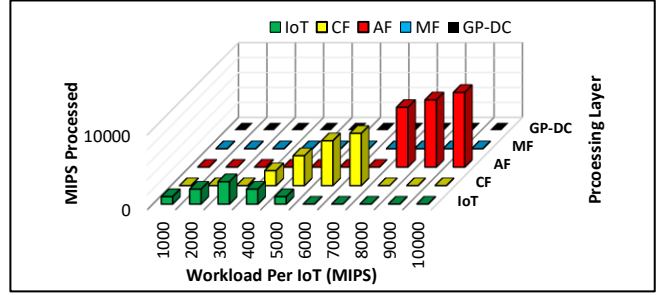

(a)

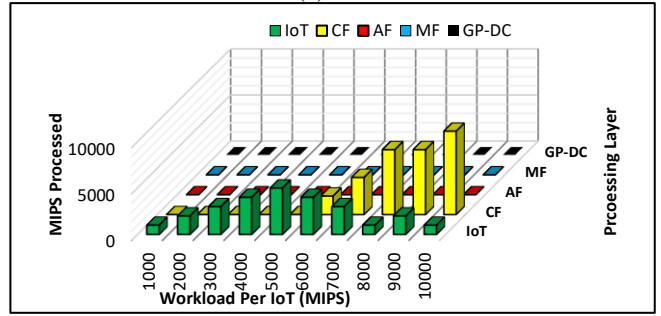

(b)

Figure 22 Workload distribution in Scenario #1 during (a) K=3 and (b) K=5 service splits, in the capacitate case.

## SCENARIO #2

In this scenario, the power savings introduced by service splitting is very limited as shown in Figure 23. This is largely due to the capacity limitations placed on the CPE fog coupled with the inflexibility posed by the restriction of service splitting. For example at 4000 MIPS, the total demand is 20,000, although the IoT devices' capacity in total can accommodate the total workload, however, this would mean the value of K has to be increased to 12, provided that 9000 MIPS was hosted at the CF layer and the remaining 11,000 MIPS was subdivided among the IoTs, hence K = 12. Interestingly, as shown in Figure 24(a) and Figure 24(b), after the total capacity of the IoT source nodes' group is depleted (4000 MIPS and beyond), the case for service splitting becomes irrelevant as the model always allocates the workload to the metro fog server as activating multiple CFs would incur high costs due to high power consumption of ONU devices.

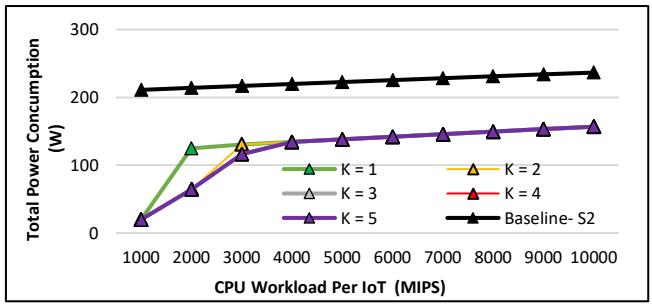

Figure 23 Total power consumption in Scenario #2 using fog, for a range of values of K, in the capacitated case.

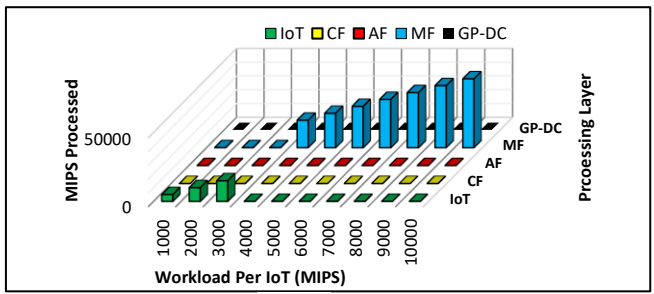
(a)

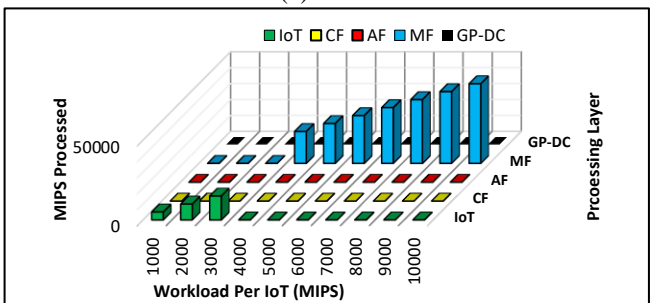
(b)

Figure 24 Workload distribution in Scenario #2 during (a) K=3 and (b) K=5 service splits, in the capacitate case.

## SCENARIO #3

The trends in this scenario remain relatively unchanged compared to Scenario #2 except for the fact that service splitting is utilized only because the IoT source nodes are from different groups, and the ONU devices would need to be turned ON anyway to get to the higher layers, hence CF attached to the ONUs are used due to their low idle power compared to the MF server. This observation was established in previous scenarios of all the cases in Figure 24, at 4000 MIPS, where only the MF server was used, compared to 4000 MIPS in this scenario where the workload is processed between the IoT and CF nodes. In this scenario, a total saving of 56% was achieved with service splitting value K >3 as opposed to 33% with no service splits K = 1, as highlighted in Figure 25. As mentioned previously, this large saving is the difference between the idle power of the MF server and the smaller devices like the ONUs and CFs. However, we have already established that, when the workload is increased, the processing per instruction at the MF compensates for the idle power of its server, hence all workloads are processed at the MF layer as can be seen in Figure 26(a) and Figure 26(b) at 7000 MIPS.

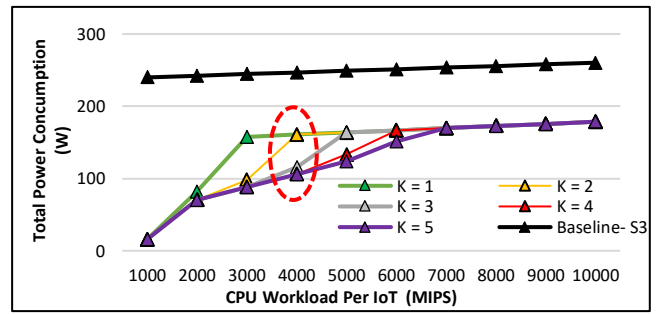

Figure 25 Total power consumption in Scenario #3 using fog, for a range of values of K, in the capacitated case.

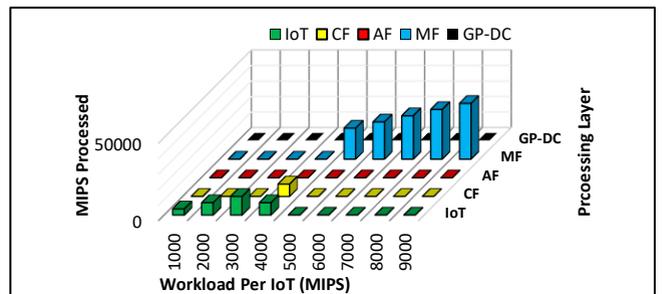
(a)

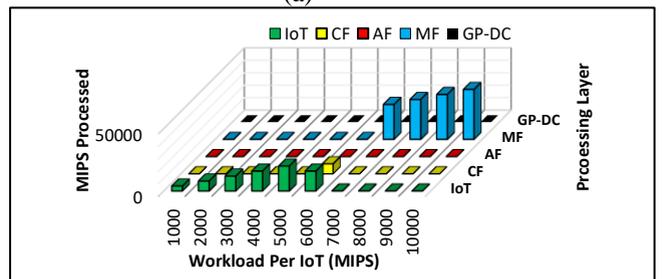
(b)

Figure 26 Workload distribution in Scenario #3 during (a) K=3 and (b) K=5 service splits, in the capacitate case.

**SCENARIO #4**

In this scenario, service splitting is predominantly irrelevant, except in rare circumstances such as the scenario at 4000 MIPS at K>1, a total of 80,000 MIPS is demanded by the source nodes and if all of this was to be processed at the MF layer, it would require two servers, hence in this case the MF server is fully packed and the remaining workload (6560 MIPS) is processed on source nodes' local CPUs. Thus, as shown in Figure 26, service splitting at K>1 introduces total savings of up to 18% compared to 0% with no service splitting (K = 1) as the solution was the same as the baseline in this instance. Similar to the observations obtained in the un-capacitated case, as shown in both cases of Figure 28, the MF and the cloud are largely the best choice, respectively, when the workload is too high.

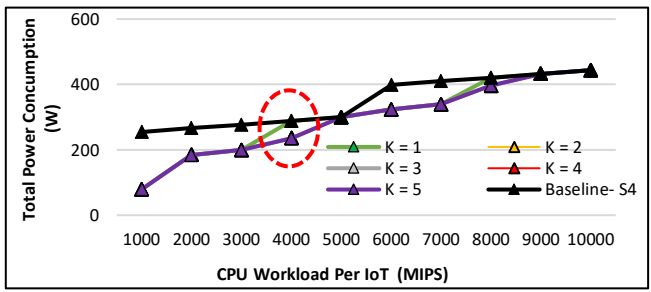

Figure 27 Total power consumption in Scenario #4 using fog, for a range of values of K, in the capacitated case.

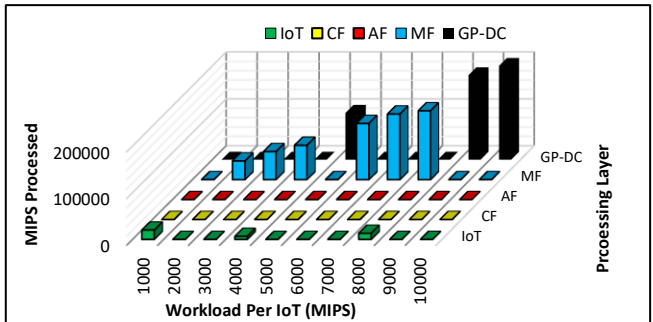

(a)

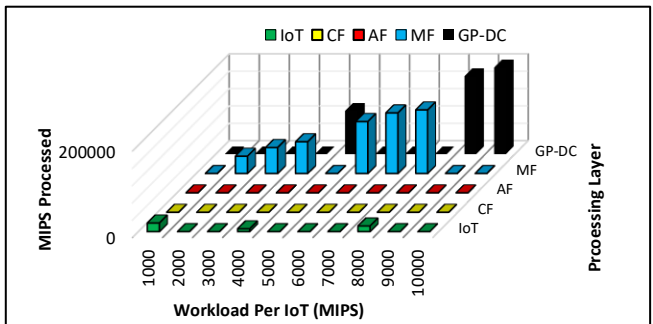

(b)

Figure 28 Workload distribution in Scenario #4 during (a) K=3 and (b) K=5 service splits, in the capacitate case.

### 4) IMPACT OF SP-DCS GIVEN NON-SPLITTABLE SERVICES

Given the high energy efficiency of SP-DC servers, it is worth investigating its impact on improving the energy efficiency of the proposed fog model and whether the fog computing approach (distributed processing) with GP-DCs is still producing savings when such highly energy efficient servers are available at the cloud, in both the capacitated and un-capacitated design problems. The results indicated that, all the trends for Scenario #1, Scenario #2 and Scenario #3, remained unchanged. However, as expected, trends changed in Scenario #4 due to the high level of workloads and the need for the cloud to intervene even prior to considering an SP-DCs, if anything, the deployment of SP-DC should incentivize centralized processing more than ever. Thus, the impact of the SP-DC is observed at and beyond 4000 MIPS, as shown in Figure 30.

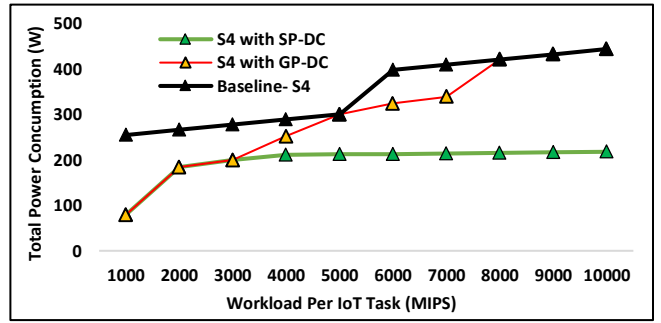

Figure 29 Total power consumption in Scenario #4 with/ without SP-DC.

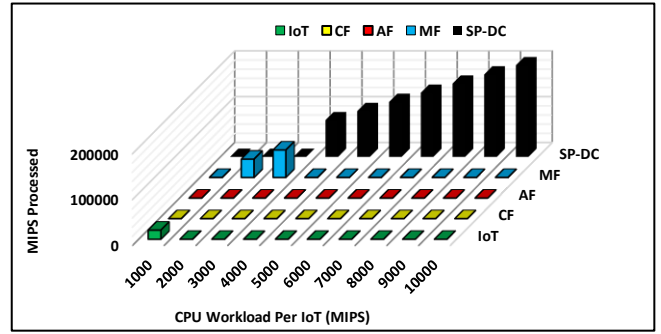

Figure 30 Workload distribution using fog, in Scenario #4 with SP-DC, in all cases.

Interestingly, as shown in Figure 29, the SP-DC yields total savings of up to 50% compared to processing in GP-DC (baseline), whilst the maximum saving obtained in the optimized scenario with GP-DCs shown in Figure 10 was up to 30%, in both capacitated and un-capacitated design problems. These results demonstrate that hosting services of high computational workload on mini DCs in the fog layers that are associated with high PUEs and are less efficient in terms of processing per instruction brings no benefits when a highly efficient centralized DC is available at the core network.

## VI. INTER-SERVICE PROCESSING OVERHEAD DUE TO SYNCHRONIZATION TRAFFIC

This section considers a scenario in which service splits incur an extra processing overhead due to synchronization between the sub-services of the service in question. We only consider processing overhead since the communication traffic power consumption is almost negligible in terms of its influence on decision making as network equipment idle power is 60% - 90% of the maximum power consumption. In future IoT networks, sub-services of an application service may need to communicate to complete a given processing task as shown in Figure 31. This communication is needed for synchronization among the sub-services, hence we have assumed a fraction of the maximum power consumption for processing (processing overhead) is accounted for at any processing server that has hosted one or multiple sub-services. As shown in the illustrative example in Figure 31, communication is established between the sub-services S1 and S2 only despite the IoT being a processing server, it does not process any sub-service(s), hence there is no need to establish any communication for synchronization in addition to S1 and S2, annotated by the green dashed lines.

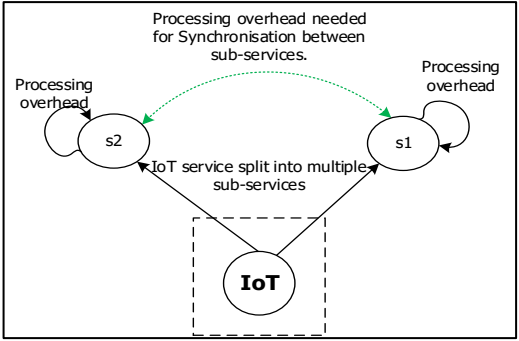

Figure 31 Illustrative example.

Before introducing the MILP model, the additional parameters and variables are defined as follows:

**Application Parameters:**
$\phi^{(p)}$  Synchronization processing overhead ratio.
$\mathbb{p}_d$  PUE of processing device $d \in P$.

**Variables:**
$\lambda^s_{d1d2}$  $\lambda^s_{d1d2} = 1$, if there is synchronisation traffic overhead of service $s \in S$ between processing node $d1 \in P$ and $d2 \in P: d2 \neq d1$, otherwise $\lambda^s_{d1d2} = 0$.
$\lambda_{d1d2}$  Synchronization traffic between processing nodes $d1 \in P$ and $d2 \in P: d2 \neq d1$.
$\lambda^{d1d2}_{mn}$  Synchronization traffic processing node $d1 \in P$ and $d2 \in P: d2 \neq d1$, traversing link $m, n$, where $m \in N$ and $n \in N_m$.
$\lambda^{(sync)}_i$  Synchronization traffic on node $i \in N$.
$\rho^s_{d1d2}$  Synchronization processing demand resulting from splitting the service request of source node $s \in S$, between processing nodes $d1 \in P$ and $d2 \in P: d2 \neq d1$.

The total power consumption equations remain intact except for an additional equation which accounts for synchronization processing overhead and this is defined as follows:

**Processing Power Consumption Overhead Due to Synchronization:**

$$\sum_{s \in S} \sum_{\substack{d2 \in P: \\ d2 \neq d1}} \left( \mathbb{p}_{d2} \rho^s_{d1d2} E^{(i)}_d \right) \quad (41)$$

**Additional Constraint:**

$$\sum_{n \in N_m} \lambda^{d1d2}_{mn} - \sum_{n \in N_m} \lambda^{d1d2}_{nm} \quad (42)$$
$$= \begin{cases} \lambda_{d1d2} & m = d1 \\ -\lambda_{d1d2} & m = d2 \\ 0 & otherwise \end{cases}$$
$$\forall d1 \in P, d2 \in P, m \in N: d1 \neq d2.$$

Constraint (42) conserves synchronization traffic from source node to destination node. It ensures that, the total incoming traffic at a node is equal to the total outgoing traffic of that node; unless the node in question is either the source node or the destination node.

$$\lambda^s_{d1d2} \leq \Omega^{sd1} \quad \forall s \in S, d1 \in P, d2 \in P: d1 \neq d2 \quad (43)$$

$$\lambda^s_{d1d2} \leq \Omega^{sd2} \quad \forall s \in S, d1 \in P, d2 \in P: d1 \neq d2 \quad (44)$$

$$\lambda^s_{d1d2} \geq (\Omega^{sd1} + \Omega^{sd2}) - 1 \quad (45)$$
$$\forall s \in S, d1 \in P, d2 \in P: d1 \neq d2$$

Constraints (43) to (45) are used in the linearization of the product of binary variables $\Omega^{sd1}$ and $\Omega^{sd2}$, where $s \in S$ $d1 \in P$ and $d2 \in P: d2 \neq d1$.

$$\rho^s_{d1d2} = \lambda^s_{d1d2} D^{(CPU)}_s \phi \quad (46)$$
$$\forall s \in S, d1 \in P, d2 \in P: d1 \neq d2$$

Constraint (46) ensures that, the total synchronization processing overhead resulting from splitting the service request of source node $s \in S$, between processing nodes $d1 \in P$ and $d2 \in P: d2 \neq d1$ is realized.

$$\lambda^{(sync)}_i = \sum_{d1 \in P} \sum_{\substack{d2 \in P: \\ d2 \neq d1}} \lambda^{d1d2}_{mn} + \sum_{d1 \in P} \sum_{\substack{n \in N_m: \\ d1 \neq m, m \in P}} \lambda^{d1m}_{nm} \quad (47)$$
$$\forall m \in N$$

Constraint (47) ensures that egress and ingress synchronization traffic on node $i \in N$ is accounted for.

$$\mathcal{N}_d \geq \frac{\left( \sum_{s \in S} \rho^{sd} + \sum_{s \in S} \sum_{\substack{d1 \in P: \\ d1 \neq d}} \rho^s_{d1d2} \right)}{C^{(CPU)}_d} \quad (48)$$
$$\forall d \in P$$

Constraint (48) determines the number of servers required at processing node $d \in P$.

$$\mathcal{N}_d \leq \mathcal{V}_d \quad \forall d \in P \quad (49)$$

Constraint (49) ensures that, the number of servers activated at a processing node $d \in P$, does not exceed the maximum available number of servers in that node.

$$\lambda_{d1d2} = \sum_{s \in S}(\lambda^s_{d1d2} D_s^{(Bw)}) \qquad (50)$$
$$\forall d1 \in P, d2 \in P: d1 \neq d2$$

Constraint (50) ensures that, the total communication demand between $d1 \in P$ and $d2 \in P$, where $d2 \neq d1$ is achieved.

### A. POWER CONSUMPTION EVALUATION

The power consumption evaluations within this section are based on the assumption that each sub-service requires 1% or 10% of the sub-service processing workload as processing overhead in order to perform synchronization among multiple sub-services of an application. Note that the number of service splits has not been constrained. The evaluation considers Scenario #1, Scenario #2 and Scenario #3 in the capacitated case, since the largest number of splits occurred in these scenarios where synchronization overhead was not accounted for. Therefore, it is of interest to investigate the extent to which the synchronization overhead impacts the decision in terms of making service splits.

### SCENARIO #1

As can be seen in previous subsection H, in Figure 22(b) Scenario #1 incurred the largest number of splits due to the geographical distribution and the nature of the service request (i.e. single demand) which incentivized splitting services onto the smaller fog nodes such as the IoT and CF devices due to their low power consumption compared to the higher layer fog nodes and the cloud. After having considered the processing overhead due to synchronization, the results in Figure 33(a), indicate that for small overhead values such as 1%, service splitting is predominantly favorable and the only time services are not split up is when the workload is very high i.e. 9000 MIPS and 10,000 MIPS. In such cases, although the model could have made use of splitting, instead it decided to consolidate the services in the AF server with much higher idle power, associated PUEs and networking overhead in return for much better processing efficiency compared to IoTs and CF servers. However, when processing overhead increased to 10%, even for very small workloads such as 2000 MIPS, the placement decision is already impacted as the services are consolidated on the CF layer as in Figure 33(b). Also in Figure 33(a), at 5000 MIPS, the original solution with no overhead (S1| No_OH) decided to split the total workload among the same group of IoTs which resulted in 5 service splits, since there was enough capacity offered by IoT devices. The current solution has done the same number of splits but due to the extra processing overhead imposed by synchronization the CF layer is used for processing. The general trend shows that, even at very small overhead ratios (e.g. 1%), service splitting in the long run is not an efficient choice as can be seen in Figure 33(a) with the increase in workload, the services are placed higher and higher up the network hierarchy so that service requests can be consolidated on fewer severs. As shown in Figure 32, power consumption is increased by up to 42%, considering 10% overhead compared to the optimization scenario with no overhead (No_OH). However, for very high workload volumes such as 10,000 MIPS, the savings are reduced by 5% since both the placement decision was not changed and the extra overhead is due to the extra processing required.

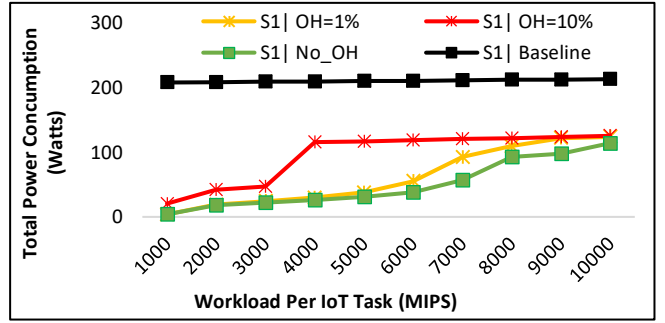

Figure 32 Power consumption of fog with/out overheads compared with the baseline solution in Scenario #1.

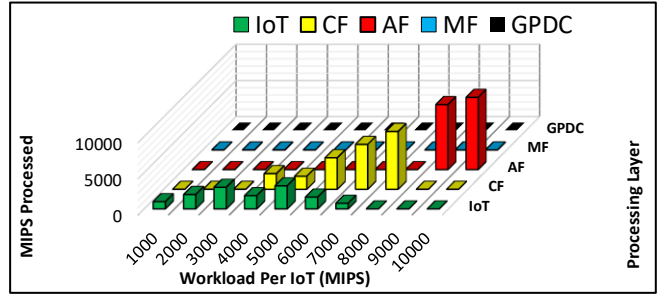

(a)

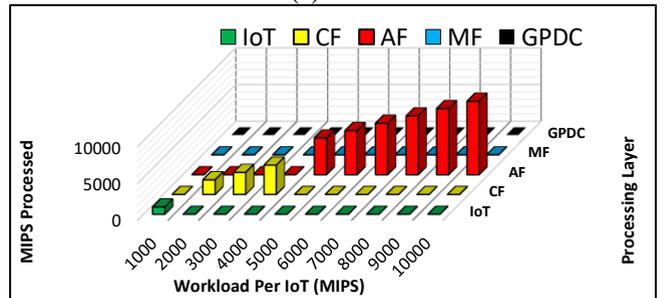

(b)

Figure 33 Workload distribution in Scenario #1 during (a) 1% CPU overhead, (b) 10% CPU overhead.

### SCENARIO #2

In this scenario, the number of active IoT's have increased, hence the total workload has also increased. As can be seen in Figure 35(a), for workloads of 2000 MIPS, the number of service splits have dropped from 10 with no processing overhead to 9, considering 1% of processing overhead. This confirms the previous observations in Scenario #1 that despite some overhead, for very low demands, service splitting does still introduce power savings which is up to 67% considering 1% overhead compared to processing in the cloud, whilst this figure drops to 42% when the overhead is increased to 10% as can be seen in Figure 34. This is mainly due to the difference between local processing on IoT devices and CF servers located in different parts of the network. Consistent with previous observations, the MF becomes the dominant choice due to its processing efficiency as this was the case prior to synchronization overhead, if anything, synchronization overhead will provide even further incentives to utilize the MF servers.

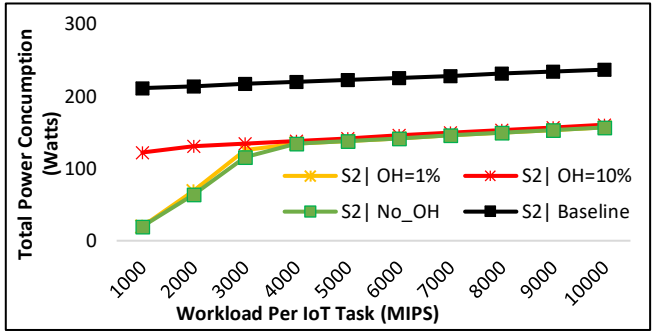

Figure 34 Power consumption of fog with/out overheads compared with the baseline solution in Scenario #2.

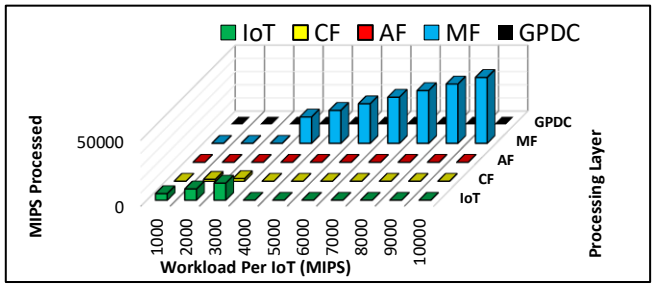

(a)

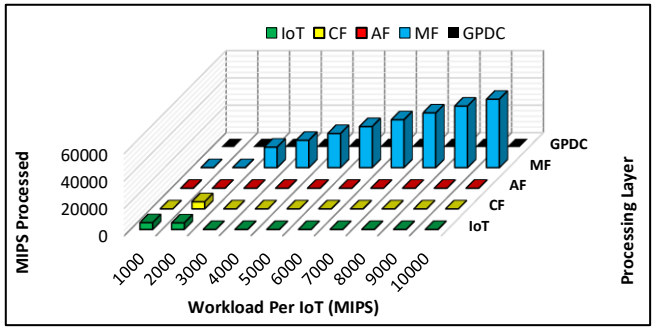

(b)

Figure 35 Workload distribution in Scenario #2 during (a) 1% CPU overhead, (b) 10% CPU overhead.

**SCENARIO #3**

In this scenario, due to the distribution of the IoT source nodes, for low workload volumes with processing overhead of 1%, service splitting is still favorable although the number of splits has decreased compared to the case with no synchronization overheads due to the intervention of the CF servers, as can be observed in Figure 37(a). In order for an IoT source node to access another IoT device to process its request, an ONU device must be activated, hence utilizing the CF servers with larger capacity would be a better packing option as it will drop number of service splits. It has already been established that, when source nodes have low service requests and that enough of idle processing resources on the IoT devices, service splitting can always produce significant savings as shown in Figure 36 at 2000 MIPS regardless of the synchronization overhead. However, when the overhead is low (i.e. 1%) and there is enough resources available on the lower fog layers (IoT and CF), the model chooses to always perform service splitting, regardless of the networking overhead incurred to access the lower fog devices in different parts of the network. Moreover, when the processing overhead is increased to 10%, service splitting is no longer the favorable choice, predominantly as seen can be seen in Figure 37(b). Total power savings drop from up to 63% with no processing overhead to up to 28% with 10% overhead as can be seen in Figure 36.

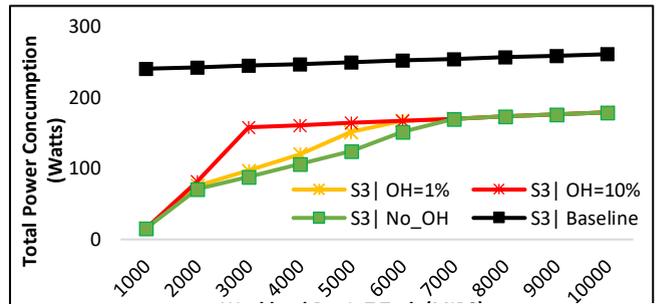

Figure 36 Power consumption of fog with/out overheads compared with the baseline solution in Scenario #3.

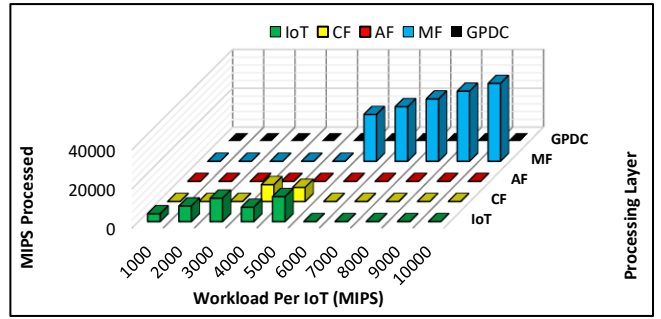

(a)

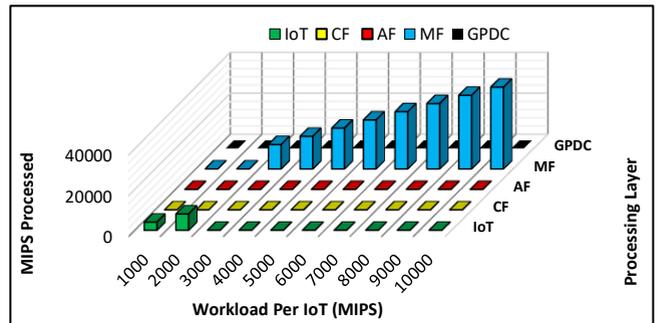

(b)

Figure 37 Workload distribution in Scenario #3 during (a) 1% CPU overhead, (b) 10% CPU overhead.

## VII. CONCLUSION

This paper has investigated the energy efficiency of distributed processing based on the concept of fog computing, for resource intensive visual-based processing services in the context of IoT. A fog architecture based on Passive Optical Networks (PONs) was proposed due to their suitability for high bit rate application services and their scalability which particularly fits well with massive IoT deployments. Several layers of processing were considered in the IoT-cloud continuum, along with emerging energy efficient special purpose data centers (SP-DCs) that are highly optimized to perform a specific service, hence, they can be much more efficient than their general-purpose counter parts (GP-DCs). We have studied two types of network design schemes; 1) capacitated and 2) un-capacitated. In the capacitated scheme, given non-splittable services and for low workload volumes, significant energy savings of up to 90% were made compared to conventional cloud due to the efficiency of local computation. However, for relatively higher workloads, the savings dropped to 30% due to activation of servers with high idle power in the metro fog (MF). As for the un-capacitated case, it was found that, generally for high workloads, building too many small CF servers was not a good option in the long run and hence the cloud DC was used due to its server processing efficiency.

We extended the work by investigating the impact of service splitting on the reduction of power consumption. It was found that in the capacitated scheme, for low workloads such as Scenario #1, a single IoT with 3000 MIPS, the savings increased from 46% with no service splitting to 88% with service splitting. However as the workload increased, service splitting was only beneficial in limited circumstances such as processing parts of the given services locally in the IoT layer in order to prevent activation of additional servers in the upper layers such as the MF and the cloud. Hence, CF and AF layers had limited or no role to play due to their high processing inefficiency and in addition to the PUE associated with AFs. The results demonstrated that, deploying SP-DCs in the cloud did not perform better than the fog except in Scenario #4 where the workload levels and the number of requests where high. It was observed that SP-DC yielded total savings of up to 50% compared to processing in GP-DC, whilst the maximum saving obtained was up to 30% using the fog approach. This is a promising performance from the SP-DC and these results demonstrate that for scenarios where the computational workload is extremely high, deploying mini DCs in the fog layers that are associated with high PUEs and are less efficient in terms of processing per instruction, hosting services on them brings no benefits when a highly efficient centralized DC is available at the core network.

Moreover, we further extended our analysis to include investigations on the impact of inter-service synchronization processing overhead between sub-services of a service request. A couple of processing overheads was used to cover two extreme cases which included 1% and 10%. The results have shown that the inter-service synchronization overhead between IoT sub-services has a great influence on the total number of service splits. The impact was evident in Scenario #1, wherein the case with no overheads, the most number of splits occurred, whereas with the lowest overhead which was 1%, the maximum number of splits reduced from 3 to 2 for the lowest ends of the demands and for the higher demands this reduced from 5 to 0 as consolidating processing in the AF server with high power consumption was more energy efficient than splitting the services among the lower layer devices such as the IoTs and CFs.

## ACKNOWLEDGMENT
The authors would like to acknowledge funding from the Engineering and Physical Sciences Research Council (EPSRC) INTERNET (EP/H040536/1), STAR (EP/K016873/1) and TOWS (EP/S016570/1) projects. The first author would like to thank EPSRC for providing his Doctoral Training Award scholarship. All data are provided in full in the results section of this paper.

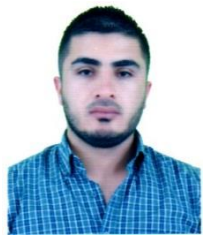

**Barzan A. Yosuf** received the BEng(Hons) with upper second-class in computer systems engineering, in 2012 and the MSc in embedded systems engineering with distinction, in 2013, from the University of Huddersfield, Huddersfield, UK. He is currently pursuing the Ph.D. degree with the School of Electronic and Electrical Engineering, University of Leeds, UK. His current research interests include IoT, fog and cloud energy optimization.

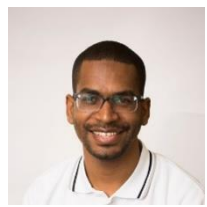

**M. Musa** received the B.Sc. (first-class Hons.) degree in electrical and electronic engineering from the University of Khartoum, Khartoum, Sudan, in 2009, the M.Sc. degree (with distinction) in broadband wireless and optical communication from the University of Leeds, Leeds, U.K., in 2011, and the Ph.D. in energy efficient network coding in optical networks from the University of Leeds, Leeds, U.K., in 2016. His current research interests include ICT energy optimization, network coding and energy efficient routing protocols in optical networks. He currently chairs working group for the IEEE P1928.1 standard, and is the secretary for the IEEE Green ICT Standards Committee.

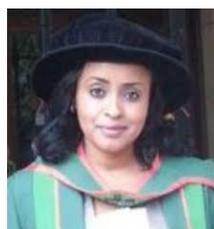

**Taisir Elgorashi** received the B.S. degree (Hons.) in electrical and electronic engineering from the University of Khartoum, Khartoum, Sudan, in 2004, the M.Sc. degree (Hons.) in photonic and communication systems from the University of Wales, Swansea, U.K, in 2005, and the Ph.D. degree in optical networking from the University of Leeds, Leeds, U.K., in 2010, where she is currently a Lecturer in optical networks with the School of Electrical and Electronic Engineering. Previously, she held a Postdoctoral research post at the University of Leeds, from 2010 to 2014, where she focused on the energy efficiency of optical networks investigating the use of renewable energy in core networks, green IP over WDM networks with datacenters, energy efficient physical topology design, the energy efficiency of content distribution networks, distributed cloud computing, network virtualization, and big data. She was a BT Research Fellow and developed energy efficient hybrid wireless optical broadband access networks and explored the dynamics of TV viewing behavior and program popularity, in 2012. The energy efficiency techniques developed during the Postdoctoral research contributed three out of eight carefully chosen core network energy efficiency improvement measures recommended by the GreenTouch consortium for every operator network worldwide. Her work led to several invited talks at GreenTouch, Bell Labs., the Optical Network Design and Modeling Conference, the Optical Fiber Communications Conference, the International Conference on Computer Communications, and the EU Future Internet Assembly in collaboration with Alcatel Lucent and Huawei.

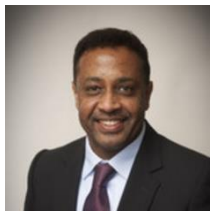

**Jaafar Elmirghani** received the Ph.D. degree in the synchronization of optical systems and optical receiver design from the University of Huddersfield, U.K., in 1994, and the D.Sc. degree in communication systems and networks from the University of Leeds, U.K., in 2014. He joined Leeds University, in 2007. Prior to that, he was the Chair in optical communications with the University of Wales, Swansea, from 2000 to 2007. He has founded, developed, and directed the Institute of Advanced Telecommunications and the Technium Digital (TD), a technology incubator/spin-off hub. He is currently the Director of the School of Electronic and Electrical Engineering, Institute of Communication and Power Networks, University of Leeds. He has provided outstanding leadership in a number of large research projects at the IAT and TD. He has coauthored Photonic Switching Technology: Systems and Networks (Wiley) and has published over 500 papers. His research interests include optical systems and networks. He is also a Fellow of the IET and the Institute of Physics. He received the IEEE Communications Society Hal Sobol Award, the IEEE Comsoc Chapter Achievement Award for excellence in chapter activities, in 2005, the University of Wales Swansea Outstanding Research Achievement Award, in 2006, the IEEE Communications Society Signal Processing and Communication Electronics Outstanding Service Award, in 2009, the Best Paper Award at the IEEE ICC'2013, the 2015 IEEE Comsoc Transmission Access and the Optical Systems Outstanding Service Award in recognition of the Leadership and Contributions to the Area of Green Communications, the GreenTouch 1000x Award, in 2015, for pioneering research contributions to the field of energy efficiency in telecommunications, the 2016 IET Optoelectronics Premium Award, and shared with six GreenTouch innovators the 2016 Edison Award in the Collective Disruption Category for their work on the GreenMeter, an international competition, clear evidence of his seminal contributions to Green Communications, which have a lasting impact on the environment (green) and society. He has been awarded in excess of the £30 million in grants to date from EPSRC, the EU, and industry and has held prestigious fellowships funded by the Royal Society and BT. He was the Co-Chair of the GreenTouch Wired, Core and Access Networks Working Group, an Adviser to the Commonwealth Scholarship Commission, and a member of the Royal Society International Joint Projects Panel and the Engineering and Physical Sciences Research Council (EPSRC) College. He was the Founding Chair of the Advanced Signal Processing for Communication Symposium, which started at the IEEE GLOBECOM'99 and has continued since at every ICC and GLOBECOM. He was also the Founding Chair of the first IEEE ICC/GLOBECOM Optical Symposium at GLOBECOM'00, the Future Photonic Network Technologies, Architectures, and Protocols Symposium. He chaired this symposium, which continues to date under different names. He was the Founding Chair of the first Green Track at the ICC/GLOBECOM at GLOBECOM 2011 and is the Co-Chair of the IEEE Sustainable ICT Initiative within the IEEE Technical Activities Board (TAB) Future Directions Committee (FDC) and within the IEEE Communications Society, a pan IEEE Societies Initiative responsible for Green and Sustainable ICT activities across IEEE, since 2012. He has been on the Technical Program Committee of 38 IEEE ICC/GLOBECOM conferences, from 1995 to 2019, including 18 times as the Symposium Chair. He was the Chairman of the IEEE Comsoc Transmission Access and Optical Systems Technical Committee and the IEEE Comsoc Signal Processing and Communications Electronics Technical Committee. He was an Editor of the IEEE COMMUNICATIONS SURVEYS AND TUTORIALS and IEEE JOURNAL ON SELECTED AREAS IN COMMUNICATIONS SERIES ON GREEN COMMUNICATIONS AND NETWORKING. He was an Editor of the IEEE Communications Magazine. He is currently an Editor of the IET Optoelectronics and Journal of Optical Communications. He was the Principal Investigator (PI) of the £6m EPSRC INTelligent Energy awaRe NETworks (INTERNET) Programme Grant (2010–2016) and is currently the PI of the £6.6m EPSRC Terabit Bidirectional Multi-user Optical Wireless System (TOWS) for 6G LiFi Programme Grant (2019–2024). He was an IEEE Comsoc Distinguished Lecturer, from 2013 to 2016.